\documentclass[reprint, superscriptaddress, amsmath,amssymb, aps,]{revtex4-2}
\usepackage[utf8]{inputenc}
\usepackage{amsmath,esint}
\usepackage{hyperref}
\hypersetup{colorlinks=true,linkcolor=blue,citecolor=blue,urlcolor=blue}
\usepackage{amsfonts}
\usepackage{amssymb}
\usepackage{bm}
\usepackage{textcomp}
\usepackage{empheq}
\usepackage{siunitx}
\usepackage[titletoc,title]{appendix}
\usepackage{graphicx}
\usepackage{dcolumn}
\usepackage{bm}
\usepackage{subcaption}
\usepackage{xcolor}
\usepackage{physics}

\usepackage{pdfpages} 
\usepackage{pgffor} 

\makeatletter
\AtBeginDocument{\let\LS@rot\@undefined}
\makeatother

\def\supplementfilename{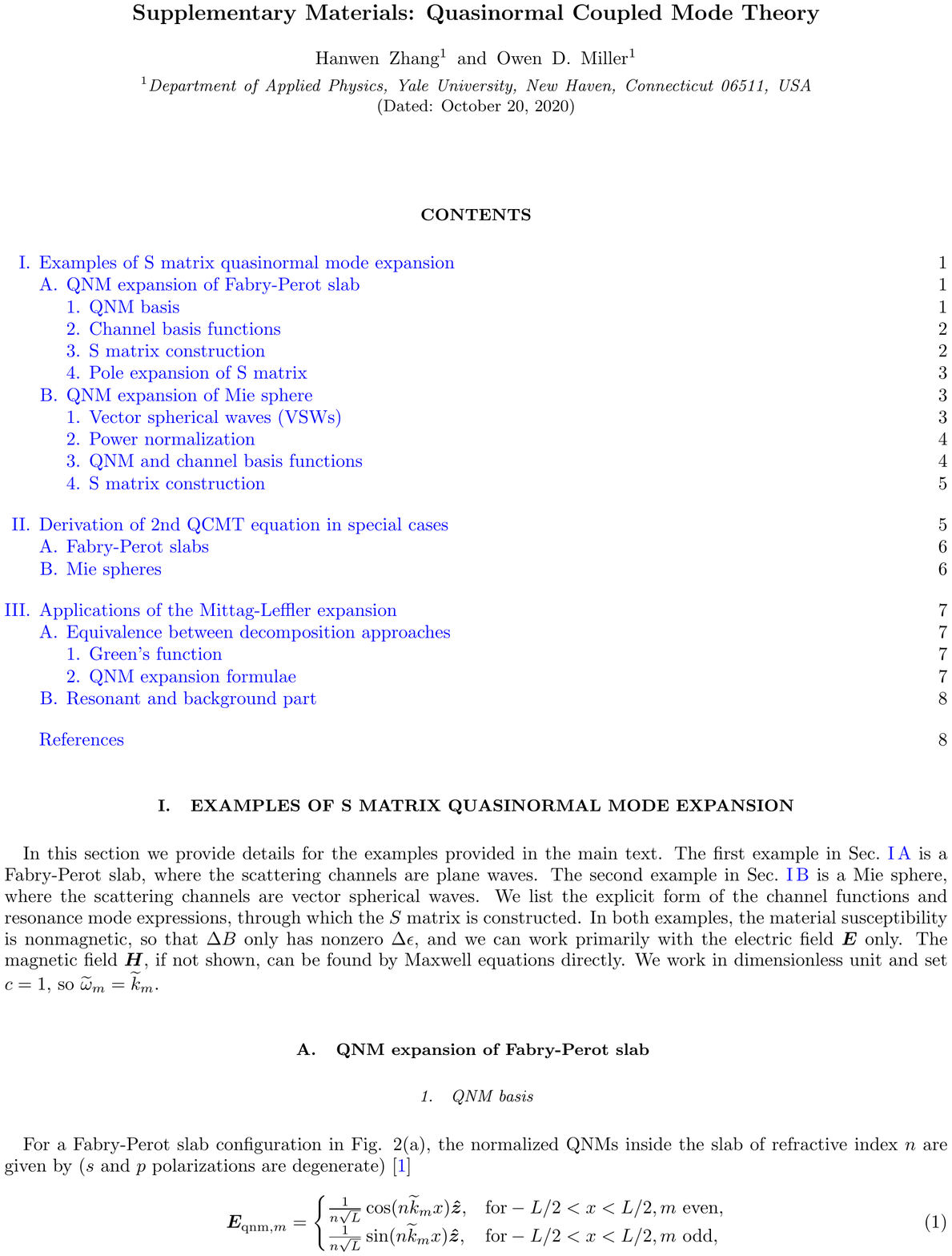}

\pdfximage{\supplementfilename}
\def\numbersupplementpages{\the\pdflastximagepages}

\newif\ifarXiv
\arXivtrue 

\renewcommand{\Im}{\operatorname{Im}}
\newcommand\I{\mathrm{i}}
\newcommand{\citeasnoun}[1]{Ref.~\cite{#1}}

\newcommand{\Figref}[1]{Fig.~\ref{fig:#1}}
\newcommand{\figref}[1]{Fig.~\ref{fig:#1}}
\renewcommand{\eqref}[1]{Eq.~(\ref{eq:#1})}
\newcommand{\Eqref}[1]{Equation~(\ref{eq:#1})}
\newcommand{\eqreftwo}[2]{Eqs.~(\ref{eq:#1},\ref{eq:#2})}
\newcommand{\eqrefrange}[2]{Eqs.~(\ref{eq:#1})--(\ref{eq:#2})}
\newcommand{\vect}[1]{\boldsymbol{\mathbf{#1}}}

\newcommand{\secref}[1]{Sec.~\ref{sec:#1}}

\newcommand*{\Sbg}{S_{\rm bg}}

\newcommand*{\SM}{{SM}}

\renewcommand{\vec}[1]{\boldsymbol{#1}}

\newcommand\D{\mathrm{d}}

\newcommand{\equals}[2]{\underbrace{#1}_{#2}}

\newcommand{\mat}[1]{\begin{matrix}#1\end{matrix}} 
\newcommand{\bb}[1]{\left(#1\right)}
\renewcommand{\sb}[1]{\left[#1\right]}

\newcommand{\cb}[1]{\left\{#1\right\}}

\newcommand\psinc{\psi_{\rm inc}}
\newcommand\pss{\psi_{\rm scat}}

\newcommand\LQNM{\Phi_{\rm Lqnm}}
\newcommand\LQNMT{\widetilde{\Phi}_{\rm Lqnm}}
\newcommand\RQNM{\Phi_{\rm Rqnm}}

\newcommand\LQNMi[1]{\psi_{\mathrm{L}, #1}}
\newcommand\RQNMi[1]{\psi_{\mathrm{R}, #1}}
\newcommand\inner[2]{\left\langle#1,#2\right\rangle}
\newcommand\out{\Phi_{\rm out}}
\renewcommand\in{\Phi_{\rm in}}
\newcommand\inc{\Phi_{\rm inc}}

\newcommand\inctr{\Phi_{\rm inc}^{\rm TR}}

\newcommand\cin{\vec{c}_{\rm in}}
\newcommand\cinc{\vec{c}_{\rm inc}}
\newcommand\cout{\vec{c}_{\rm out}}
\newcommand\titwi[1]{\widetilde{\omega}_{#1}}

\newcommand\tit[1]{\widetilde{#1}}
\begin{document}
\preprint{APS/123-QED}

\title{Quasinormal Coupled Mode Theory}

\author{Hanwen Zhang}
\affiliation{Department of Applied Physics, Yale University, New Haven, Connecticut 06511, USA}
\author{Owen D. Miller}
\affiliation{Department of Applied Physics, Yale University, New Haven, Connecticut 06511, USA}

\date{\today}
\begin{abstract}
    Coupled mode theory (CMT) is a powerful framework for decomposing interactions between electromagnetic waves and scattering bodies into resonances and their couplings with power-carrying channels. It has widespread use in few-resonance, weakly coupled resonator systems across nanophotonics, but cannot be applied to the complex scatterers of emerging importance. We use quasinormal modes to develop an exact, \emph{ab initio} generalized coupled mode theory from Maxwell's equations. This quasinormal coupled mode theory, which we denote ``QCMT'', enables a direct, mode-based construction of scattering matrices without resorting to external solvers or data. We consider canonical scattering bodies, for which we show that a CMT model will necessarily be highly inaccurate, whereas QCMT exhibits near-perfect accuracy.
\end{abstract}

\pacs{Valid PACS appear here}

\maketitle

\section{Introduction}
Coupled mode theory (CMT) is an indispensable theoretical tool for decomposing complex electromagnetic scattering problems into the interaction of power-carrying channels with resonant modes~\cite{haus1984waves,joannopoulosphotonic,suh2004temporal}. Yet CMT requires high-quality-factor modes in a weak-coupling limit~\cite{joannopoulosphotonic}, preventing quantitatively or even qualitatively accurate predictions for lossy or non-Hermitian photonic materials, complex multi-resonance metagrating and metasurface structures, or resonators with high radiative-loss rates. In this Article, we show that with only the assumption of a basis of quasinormal modes (QNMs) within scattering bodies~\cite{leung1994completeness,lalanne2018light}, one can map \emph{any} electromagnetic scattering problem to a generalized CMT model, without restrictive frequency, quality-factor, or weak-coupling requirements. Our quasinormal coupled mode theory (QCMT) comprises exact and intuitive coupling matrices, enabling first-principles modal calculations of the full scattering matrix $S$ of any system. We use a Mittag--Leffler expansion to our $S$ matrix to simplify the frequency dependencies of the coupling matrices, cleanly separating the ``resonant'' and ``background'' contributions to the scattering process. We demonstrate a unification and equivalence of a few different QNM expansion formulae that have been proposed. We show that conventional conservation laws such as $K^\dagger K = 2\Im \Omega$, where $K$ is a mode--channel coupling matrix and $\Omega$ comprises the resonant frequencies, do not necessarily apply in general scattering problems, though we identify the limits in which they are recovered. Our work unites two important frameworks that have developed in parallel---temporal coupled-mode theory, typically for high-quality factor, isolated resonances, and quasinormal modes, typically for low-quality-factor and overlapping resonances.

\subsection{Background}
Coupled mode theory~\cite{haus1984waves,joannopoulosphotonic,fan2003temporal,suh2004temporal} offers a prescription for decomposing complex scattering problems into \emph{channels}, that carry power into and out of the system, and \emph{modes}, which may control the response of the scatterer. For applications ranging from waveguide filters~\cite{manolatou1999coupling} to photovoltaic absorbers~\cite{yu2010fundamental} to transparent displays~\cite{hsu2014transparent}, CMT enables predictions of high-performance designs and fundamental limits. Yet such predictions are only valid within the limits of CMT itself, which requires a key assumption: weak coupling between high-quality-factor ($Q$) resonances~\cite{joannopoulosphotonic}. This assumption is violated in many scenarios of emerging interest (including photovoltaic absorbers). For example, plasmonic structures~\cite{khurgin2015deal,ishii2016titanium} have significant material losses, large-area metasurfaces~\cite{lalanne2017metalenses,khorasaninejad2017metalenses,banerji2019imaging} comprise large numbers of low-$Q$, highly coupled resonances, and random media that are encountered in wavefront-shaping applications~\cite{hsu2017correlation} have significant nonresonant contributions to their response. Clearly there is a need for a CMT-like framework without the assumption of weakly-coupled high-$Q$ resonances.

An emerging technique for analyzing the modal structure of a scattered field is to decompose it into \emph{quasinormal} modes~\cite{lalanne2018light,sauvan2013theory,muljarov2016resonant}, which are eigenfields of the generally non-Hermitian Maxwell operator. There are multiple approaches to such QNM expansions, e.g., via orthogonality decompositions~\cite{sauvan2013theory,yan2018rigorous} or complex-analysis-based Mittag--Leffler expansions~\cite{muljarov2016resonant,muljarov2016exact}. These techniques have been used to successfully apply modal analysis to plasmonics~\cite{sauvan2013theory,yan2018rigorous} and diffraction gratings~\cite{gras2019quasinormal}, where a normal-mode approximation (as used in CMT) would necessarily be inaccurate. Yet while these approaches can decompose scattered fields into modes, they have not successfully captured the full interactions of the incident and scattered waves with the incoming and outgoing channels, which is the second key component to a CMT-like theory. Consequently, none of the previous approaches yield all relevant CMT equations, nor can they construct the full scattering matrices of the system. References~\cite{alpeggiani2017quasinormal,weiss2018calculate} construct scattering matrices but require fitting procedures and alternative full-Maxwell solvers to do so.

There is an alternative construction of scattering matrices with a CMT-like appearance that partitions open systems into a closed compact cavity, with discrete spectrum, and an open exterior, with a continuous spectrum. In this approach, which originated in nuclear scattering theory~\cite{mahaux1969shell,beenakker1997random,viviescas2003field}, the quasinormal modes comprise only the cavity modes (in distinction with the nanophotonics convention, which includes, e.g., ``PML'' modes~\cite{yan2018rigorous}), and a self-energy operator couples the interior to the exterior. The resulting nonlinear matrix expressions can be quite difficult to compute and are more frequently used for their analytical properties~\cite{sweeney2019theory,rotter2017light}. By contrast, the QCMT expressions we derive can be computed by standard Maxwell solvers~\cite{lalanne2019quasinormal}.

In this work, we develop a complete CMT-like theory with quasinormal modes. Conventional CMT, briefly summarized in \secref{CMT}, comprises two linear matrix equations, the first of which connects the excited mode amplitudes to the incoming-wave amplitudes, and the second of which connects the outgoing-wave amplitudes to the incoming waves and the resonant excitations. The key result of our paper is the derivation of a set of two analogous equations based on quasinormal modes. To derive these equations, one first needs a formal description of both QNMs and scattering channels, which we provide in \secref{framework}. We then derive the two key QCMT equations through integral-equation identities that connect the fields external to the scatterer to fields within the scatterer, where they can be expanded in QNMs. The first equation has appeared in various (not obviously identical) forms in previous works~\cite{sauvan2013theory,yan2018rigorous,muljarov2016resonant,lalanne2018light}, whereas the second QCMT equation has to our knowledge not appeared previously. We use these two equations to derive the QNM-based full scattering matrix of any system (\secref{deriv}), which comprises frequency-dependent coupling matrices (unlike their frequency-independent conventional counterparts). Our key results are summarized in \secref{QCMT}, representing the foundational components of QCMT. In \secref{pole}, we show that Mittag--Leffler pole expansions of our derived expressions, when certain asymptotic conditions are satisfied, can lead to simplifications of the coupling matrices. We consider two canonical scattering problems as test examples of QCMT: one-dimensional Fabry-Perot scattering, and three-dimensional Mie scattering, where we demonstrate the accuracy of QCMT, and the necessary \emph{inaccuracy} of conventional CMT (\secref{examples}). A more general discussion of the valid regimes of conventional CMT follows in \secref{cmt_acc}; interestingly, we find that QCMT shows a similar time-domain equation structure as its conventional CMT counterpart, except that there is an additional direct-scattering term and all coupling matrices are convolution operators in time, arising from the inherent frequency dependencies of the underlying Maxwell operators.

\subsection{Conventional coupled mode theory (CMT)}
\label{sec:CMT}
Conventional coupled mode theory has been applied extensively to resonant phenomena across the nanophotonics landscape~\cite{joannopoulosphotonic,fan2002analysis,fan2003temporal,yanik2003all,hamam2007coupled,verslegers2010temporal,verslegers2012electromagnetically,ruan2010superscattering,ruan2012temporal,yu2010fundamental,hsu2014theoretical} . There are four key matrices that appear in coupled-mode theory. The first, $\Omega$, dictates the phase and amplitude evolution of modal amplitudes $\vec{a}$. Its entrywise real parts determine the phase evolution, while its entrywise imaginary parts are amplitude decay rates. Incoming waves with coefficients $\cin$ couple to the modal amplitudes $\vec{a}$ by the transpose of a matrix $D$, while the reverse process of modal amplitudes coupling to outgoing waves is described by a matrix $K$. Finally, the ``direct'' scattering process of incoming waves coupling directly to outgoing waves is captured in a matrix $\Sbg$. Often this matrix is denoted as a ``$C$'' matrix~\cite{suh2004temporal}, but we use $S_{\rm bg}$ because it physically represents the background scattering matrix. These relations are summarized in the CMT equations, which at frequency $\omega$ ($e^{-i\omega t}$ harmonic time dependence) read~\cite{suh2004temporal}:
\begin{align}
    \I(\Omega - \omega)\vec{a} &= D^{T}\cin \label{eq:cmt1} \\
    \cout &= \Sbg\cin  + K\vec{a} \label{eq:cmt2}\,,
\end{align}
which are derived from an ansatz of weak coupling. Importantly, the coupling matrices are frequency-independent, which is not true in the general QCMT framework we derive in the following sections. The coupling matrices must satisfy certain constraints~\cite{suh2004temporal}:
\begin{align}
    K &= D \qquad \textrm{(reciprocal)} \label{eq:KD} \\
    K^\dagger K &= -2 \Im \Omega \label{eq:KdagK} \\
    \Sbg D^* &= -D, \label{eq:Dbg}
\end{align}
where for the matrix $\Omega$, the notation $\Im \Omega$ is its anti-Hermitian part, i.e., $\Im \Omega = (\Omega - \Omega^\dagger)/2\I$.  The first relation is required only in reciprocal systems, within which the coupling of incoming waves to modes is the reciprocal scenario of modes coupling to outgoing waves. (More generally, generalized reciprocity~\cite{Kong1975} would enforce a condition between $K$ of the given system and $D$ for a material-transposed system~\cite{Mann2019}.)  The second relation is a consequence of energy conservation: the total decay rate of all modes must equal the power flowing from the modes to the outgoing channels. The final relation arises from the requirement of no outgoing waves in the presence of an input that is exactly matched to the resonant mode via time reversal.

By substituting for the modal amplitudes in \eqreftwo{cmt1}{cmt2}, one can directly determine the scattering matrix $S$ that relates the outgoing-wave coefficients $\cout$ to the incoming-wave amplitudes $\cin$:
\begin{equation}
    S = \Sbg -\I K (\Omega - \omega)^{-1} D^T.
    \label{eq:cmt_smat}
\end{equation}
Typically, the CMT equations are used for semi-analytical descriptions with a single resonance~\cite{joannopoulosphotonic,fan2003temporal,hamam2007coupled}, or a small number of isolated resonances~\cite{suh2004temporal,hsu2014theoretical}, in which cases \eqrefrange{cmt1}{cmt_smat} can provide accurate descriptions. Yet for scattering bodies that exhibit any complexity in their resonant response, even high-symmetry scatterers like a Mie sphere (as discussed in \secref{examples}), these equations fail to accurately capture scattering response, and a more general representation is required.

\begin{figure}[h]
    \centering
    \includegraphics[width=0.8\linewidth]{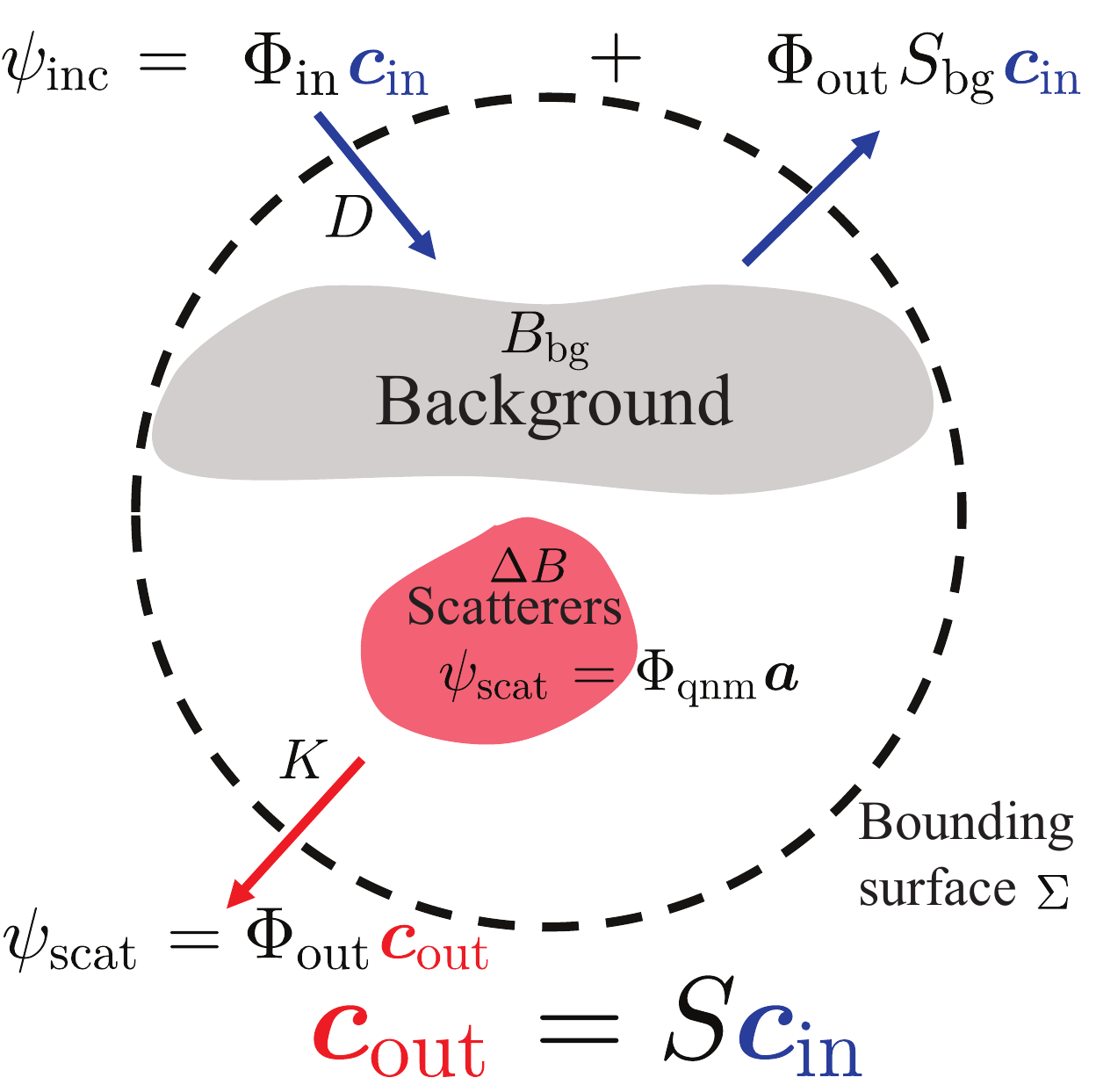}
    \caption{An incident field $\psinc$ in a background of material tensor $B_{\rm bg}$ impinges upon scatterers, with susceptibility $\Delta B$, exciting a scattered field $\pss$. The coupling of the incident field to the resonances is described by a coupling matrix $D(\omega)$ in a relation that comprises the first QCMT equation, \eqref{cqmt1}. Within the scatterers, the scattered field can be decomposed into quasinormal modes $\Phi_{\rm qnm}$, with modal amplitues $\vec{a}$, which leak out into radiation amplitudes determined by a coupling matrix $K(\omega)$, as dictated by the second QCMT equation, \eqref{cqmt2}. The ``channels'' carrying power into and out of the scattering bodies are defined on a bounding surface $\Sigma$, where they satisfy power-orthonormality relations, \eqref{power_normal}. The $S$-matrix connecting the incoming-wave coefficients, $\cin$, to the outgoing-wave coefficients, $\cout$, is determined by the QCMT equation of \eqref{cqmt_smat}.}
    \label{fig:scat_frame}
\end{figure}

\section{Scattering Framework}
\label{sec:framework}
In this Section we introduce the basic theoretical techniques and terminology for developing the quasinormal coupled mode theory. First we describe the separation of Maxwell fields into incident and scattered fields, and the governing equations relating them, and then we describe a formulation of orthogonal ``scattering channels'' that carry power and momentum into and/or out of a scattering body (\secref{channels})~\cite{liu2018optimal}.
\subsection{Maxwell fields}
\label{sec:fields}
We write Maxwell's equations in compact six-vector notation
\begin{equation}
	\equals{\bb{\mat{&-\curl\\-\curl &\\}}}{\Theta}\equals{\bb{\mat{\vec{E}\\\vec{H}}}}{\psi} - \I\omega\equals{\bb{\mat{\epsilon & \zeta\\ -\eta & -\mu}}}{B}\bb{\mat{\vec{E}\\\vec{H}}}  = - \equals{\bb{\mat{\vec{J}_e\\-\vec{J}_m}}}{\xi}
	\label{eq:maxwell}
\end{equation}
where we allow for arbitrary material anisotropy (or bianisotropy) and do not assume reciprocal materials. We include negative signs in the magnetic-field terms such that the 6$\times$6 Green's tensor is complex-symmetric when the materials are reciprocal. (Without the negative signs, reciprocity manifests in a more complicated symmetry condition including sign-flip matrices~\cite{miller2013photonic}, while representing the same physics.) Writing the curl operator as $\Theta$, the 6$\times$6 material tensor as $B$, the 6-component field tensor as $\psi$, and source currents as $\xi$,the six-vector version of Maxwell's equations is:
\begin{equation}
	\Theta \psi - \I\omega B \psi =  - \xi\,.
	\label{eq:short_mw}
\end{equation}
The material tensor $B$ is frequency-dependent for dispersive materials.

\subsection{Scattering channels}
\label{sec:channels}
Scattering channels carry power towards and away from a scatterer. They are 6-component vector fields comprising bases $\out$ and $\in$:
\begin{equation}
	\out = \bb{\mat{\vec{E}_{\rm out,1} & \vec{E}_{\rm out,2} & \ldots\\\vec{H}_{\rm out,1} & \vec{H}_{\rm out ,2} & \ldots}}\,, 	\in = \bb{\mat{\vec{E}_{\rm in,1} & \vec{E}_{\rm in,2} & \ldots\\\vec{H}_{\rm in,1} & \vec{H}_{\rm in,2} & \ldots}}\,,
	\label{eq:out_in}
\end{equation}
with their key trait being that they satisfy a power orthonormality condition on a bounding surface, source/receiver volumes, or analogous domains~\cite{Miller2019}. In six-vector notation the Poynting flux of a field $\psi$ through a surface $\Sigma$ can be written $\int_{\Sigma} \psi^\dagger Q \psi / 4$, where $Q = \bb{\mat{& -\vec{n}\times\\ \vec{n}\times &\\}}$ is Hermitian and $\vec{n}$ is the outward surface normal vector. We assume the basis fields are normalized to have unity outgoing and incoming powers, with an inner product defined by the Poynting flux. Then the inner products of the basis functions satisfy
\begin{align}
    \inner{\out}{\out} &= \int_{\Sigma} \out^{\dagger}\frac{Q}{4}\out \,{\rm d}S = \mathbb{I},\nonumber\\
    \inner{\in}{\in} &= \int_{\Sigma} \in^{\dagger}\frac{Q}{4}\in \,{\rm d}S = -\mathbb{I}\nonumber\\
    \inner{\in}{\out} &=  \int_{\Sigma} \in^{\dagger}\frac{Q}{4}\out =  0\,,
	\label{eq:power_normal}
\end{align}
where $\mathbb{I}$ represents the identity matrix. Typical examples of scattering channels include vector spherical waves~\cite{bohren2008absorption,kristensson2016scattering}, plane waves~\cite{verslegers2012electromagnetically,zhang2019scattering,chung2020high}, and the singular vectors of the scattering matrix of a given scatterer~\cite{miller2017universal,Miller2019}. In many cases the incident wave is a traveling wave, such as a plane wave, which is a combination of both incoming and outgoing fields, so it is useful to define an incident-wave basis $\inc$ comprising both incoming and outgoing waves,
\begin{align}
    \inc = \alpha\in + \beta\out,
    \label{eq:incbasis}
\end{align}
where $\alpha$ and $\beta$ are constants. Up to arbitrary phase factors, for vector spherical waves, $\alpha = \beta = \frac{1}{2}$; for plane waves, 
 $\alpha = \beta = 1$. (See {\SM} for details of the plane wave basis definition.)
Any given incident wave can be represented in the incoming/outgoing-wave basis by the expression
\begin{equation}
	\psinc = \inc\cinc = \in\cin + \out \Sbg\cin.
	\label{eq:incs}
\end{equation}
By matching the vectors multiplying $\in$ in \eqreftwo{incbasis}{incs}, we can see that $\cinc = \frac{1}{\alpha}\cin$.


\subsection{Quasinormal modes}
Quasinormal modes are eigenfields of a non-Hermitian Maxwell operator, where non-Hermiticity arises from material loss, radiation loss (open boundaries), linear gain, or any other source of power imbalance in a resonator. Whereas \emph{normal} modes are solutions of Hermitian eigenproblems and are guaranteed to form a complete basis for the fields in a scatterer, the solutions of a non-Hermitian eigenproblem do not necessarily form a complete basis, and these singularities are the ``exceptional points'' that have garnered tremendous recent interest~\cite{liertzer2012pump,hodaei2017enhanced,pick2017general,miri2019exceptional}. Yet these are, necessarily, singular points in the space of all possible field and material distributions, and at any generic perturbation away from an exceptional point the quasinormal modes will comprise a complete basis~\cite{pick2017general}. In fact, QNMs are likely to be over-complete~\cite{leung1994completeness,hoenders2006quasi}---only a subset are likely necessary to form the expansion basis---and this leads to multiple equivalent expansion formulae and various sum rule relations~\cite{muljarov2016resonant}.


Since we do not assume reciprocity, for any eigenfrequency $\titwi{m}$, indexed by $m$, we have to distinguish between right quasinormal modes $\RQNMi{m}$ and left quasinormal mode $\LQNMi{m}$:
\begin{align}
    \Theta\RQNMi{m} &= \I\titwi{m}B(\titwi{m})\RQNMi{m}\,, \label{eq:rqnm} \\
    \Theta \LQNMi{n} &= \I\titwi{n}B^T(\titwi{n})\LQNMi{n}\,. \label{eq:lqnm}
\end{align}
where the transpose in the material $B$ operator arises from $\LQNMi{n}$ being defined on the left of the operators $\Theta$ and $B$, transposing the entire equation, and utilizing the complex-symmetry of the $\Theta$ operator ($\Theta^T = \Theta$). For reciprocal materials, $B=B^T$ and the left QNMs (comprising the ``dual basis'' ~\cite{pick2017general}) coincide with the right QNMs. We assume any standard computational discretization of the problem (with sufficiently high accuracy)~\cite{jin2011theory}, and write the QNMs as columns in a basis matrix:
\begin{equation}
	\RQNM = \bb{\mat{\RQNMi{1} & \RQNMi{2} & \ldots}} \,,
	\label{eq:rqnm_mat}
\end{equation}
with corresponding eigenvalues
    \begin{equation}
        \Omega = \bb{\mat{\widetilde{\omega}_1 & & & \\ & \widetilde{\omega}_2\\ & &\ddots}}\,,
	\label{eq:rqnm_eig}
\end{equation}
and similarly define $\LQNM$ for the left QNMs, which share the same eigenvalues. One can multiply \eqref{rqnm} and \eqref{lqnm} on the left by $\LQNMi{n}^T$ and $\RQNMi{m}^T$, respectively; integrating over all space $V$ yields the QNM orthogonality relation~\cite{sauvan2013theory}:
\begin{equation}
	\int_V \LQNMi{n}^T\bb{\titwi{n}B(\titwi{n}) - \titwi{m}B(\titwi{m})}\RQNMi{m} = 0\,.
	\label{eq:qnm_ortho}
\end{equation}
To normalize the individual QNMs, we consider \eqref{qnm_ortho} with the left and right QNMs indexed by the same value $n$; the material-dependent term in the middle goes to zero, but instead one can divide by $\titwi{n} - \omega$ and take the limit as the difference goes to zero:
\begin{eqnarray}
    &&\lim_{\omega\rightarrow\titwi{n}}\int_V\LQNMi{n}^T\bb{\frac{\titwi{n}B(\titwi{n}) - \omega B(\omega)}{\titwi{n}-\omega}}\RQNMi{n}\nonumber\\
	&=&\int_V\LQNMi{n}^T \frac{\partial }{\partial \omega}\sb{\omega B(\omega)}_{\omega=\titwi{n}}\RQNMi{n}\, = 1.
	\label{eq:normalization}
\end{eqnarray}
\Eqref{normalization} cannot be used to independently normalize both $\LQNMi{n}$ and $\RQNMi{n}$ for a given $n$, but in any scattering computation one only needs overlap integrals between them, for which \eqref{normalization} is sufficient. For nondispersive materials, \eqref{qnm_ortho} reduces to an orthogonality relation
\begin{equation}
	\bb{\titwi{n}- \titwi{m}}\int_V \LQNMi{n}^T B \RQNMi{m} = 0\,,
	\label{eq:qnm_ortho_nd}
\end{equation}
and the normalization is simply $\int_V \LQNMi{n}^T B \RQNMi{n} = 1$. Once all relevant modes are solved for, they become the natural basis for the fields inside of scatterers.

Two recent works attempt to construct the full scattering matrix directly from quasinormal modes. Reference~\cite{alpeggiani2017quasinormal} does so, albeit with unknown background-scattering-term parameters that are computed by (incorrectly) assuming the validity of conventional CMT. Reference~\cite{weiss2018calculate} constructs the resonant part of the scattering matrix from QNMs, but misses a direct-scattering term, yielding an incomplete construction. It appears that the key expression missing from these works is an equation deriving outgoing-wave coefficients from QNM amplitudes and incoming-wave coefficients, which we determine in our ``QCMT equation 2'' in \eqreftwo{cqmt_second}{cqmt2} below.

\section{From Maxwell to QCMT}
\label{sec:deriv}
A coupled-mode-theory representation decomposes a scattering problem into three sets of degrees of freedom: the incoming-wave amplitudes, the outgoing-wave amplitudes, and the amplitudes of the resonances, i.e., the quasinormal modes. The incoming-wave amplitudes are specified by the problem of interest. This leaves two sets of amplitudes to solve for: the QNM amplitudes (\secref{eqn1}) and the outgoing-wave amplitudes (\secref{eqn2}). We derive the two corresponding QCMT equations in the following subsections.
\subsection{QCMT Eqn 1: QNM amplitudes}
\label{sec:eqn1}
The first QCMT equation should relate the quasinormal-mode amplitudes, $\vec{a}$, to the incoming-wave amplitudes $\cin$ (like its CMT counterpart, \eqref{cmt1}). To do this we need to find the QNM response for a given incident field. In the region of the scatterer, the scattered field $\pss$ can be decomposed into the QNMs:
\begin{equation}
	\pss =
        \begin{pmatrix}
            & & \\
            \RQNMi{1} & \RQNMi{2} & \cdots \\
            & & \\
        \end{pmatrix}
        \begin{pmatrix}
            a_1 \\
            a_2 \\
            \vdots
        \end{pmatrix}
        =
        \RQNM\vec{a},
	\label{eq:rqnm_exp}
\end{equation}
where $\RQNM$ is the basis of QNM resonance fields and $\vec{a}$ are the unknown expansion coefficients. To relate the expansion coefficients to the incoming-wave amplitudes, it might be possible to use the typical differential form of Maxwell's equations, but the fields have independent degrees of freedom everywhere in space, including outside the scatterer. Instead, all of the degrees of freedom can be brought inside the scatterer by using the volume equivalence principle~\cite{jin2011theory}. The material tensor $B$ can be separated into background and scatterer constituents,
\begin{equation}
	B = B_{\rm bg} + \Delta B\,.
	\label{eq:inc_scat}
\end{equation}
The incident field $\psi_{\rm inc}$ is the solution of Maxwell's equations in the absence of the scatterer (i.e. with $B =  B_{\rm bg}$ everywhere), while the scattered field $\pss$ is the additional field excited when $\Delta B$ is introduced, which is given by the difference between the total and incident fields, $\pss = \psi - \psi_{\rm inc}$. Then straightforward manipulation of Maxwell's equations yields a differential equation for $\psi_{\rm scatt}$ with all degrees of freedom within the scatterer~\cite{jin2011theory}:
\begin{equation}
    \Theta \psi_{\rm scatt} - \I\omega B\psi_{\rm scatt} = \I\omega\Delta B \psi_{\rm inc}\,, \label{eq:v_eqv}
\end{equation}
One can see that \eqref{v_eqv} relates $\pss$ to $\psi_{\rm inc}$, and thus should also define the connection from $\vec{a}$ to $\cin$. Inserting the expression relating $\psi_{\rm inc}$ to $\cin$, \eqref{incbasis}, into the right-hand side of \eqref{v_eqv}, and the decomposition of $\pss$ into QNMs, \eqref{rqnm_exp}, on the left-hand side, and utilizing generalized reciprocity relations as well as the normalization properties of the QNMs, \eqreftwo{qnm_ortho}{normalization}, leads directly to the first QCMT equation:
\begin{equation}
	N(\omega)\I(\Omega-\omega)\vec{a} = \frac{\I\omega}{\alpha}(\LQNM, \Delta B \inc)\cin \,,
	\label{eq:cqmt_first}
\end{equation}
where  $N(\omega)$ is a nonsingular matrix  with entries $N_{nm}(\omega) = \int_V\LQNMi{n}^T\bb{\frac{\titwi{n}B(\titwi{n}) - \omega B(\omega)}{\titwi{m}-\omega}}\RQNMi{m}$. If $B$ is nondispersive, the normalization condition of \eqref{normalization} implies that $N$ is the identity matrix.

\Eqref{cqmt_first} is our first QCMT equation. It is not new, having been derived in Refs.~\cite{sauvan2013theory,lalanne2018light}, but we included a brief derivation for completeness. The intuition behind \eqref{cqmt_first} is similar to that of the first CMT equation, \eqref{cmt1}: the QNM amplitudes $\vec{a}$ are large for the modes (within $\LQNM$) that have large overlap with the incident field over the scatterer volume, and/or whose resonant frequencies have small imaginary parts and real parts close to the excitation frequency $\omega$. The matrix $N(\omega)$ accounts for modal overlaps in dispersive media, though as we show in \secref{pole}, this matrix can be dropped from the equation when a Mittag--Leffler expansion is valid.

\subsection{QCMT Eqn 2: Outgoing-channel wave amplitudes}
\label{sec:eqn2}
The second QCMT equation, analogous to \eqref{cmt2} of conventional CMT, should determine the outgoing-wave amplitudes from the incoming-wave and quasinormal-mode amplitudes. First, we recognize that the outgoing-wave amplitudes are given by an overlap integral on the bounding region $\Sigma$ of the outgoing-wave basis functions with the total field $\psi$:
\begin{align}
    \cout = \inner{\out}{\psi}= \int_{\Sigma}\out^{\dagger}\frac{Q}{4}\psi.
    \label{eq:cout_def}
\end{align}
We can write the total field as the incident field plus the scattered field, with the scattered field itself being the field radiated from the polarization field inside the scatterer via a convolution with the background Green's function, $\Gamma_{\rm bg}$:
\begin{align}
    \pss(\vec{r}) = \I\omega\int_{V'}\Gamma_{\rm bg}(\vec{r},\vec{r}')\Delta B(\vec{r}')\psi(\vec{r}').
    \label{eq:v_int_eq}
\end{align}
The polarization field itself must be separated into incident and scattered fields, the latter of which comprise the QNMs, per \eqref{rqnm_exp}. Making these substitutions, we have
\begin{align}
    \cout &=\int_{\Sigma}\out^{\dagger}\frac{Q}{4}\psinc(\vec{r}) \nonumber \\
          &+ \I\omega\int_{\Sigma}\out^{\dagger}\frac{Q}{4}\int_{V}\Gamma_{\rm bg}(\vec{r},\vec{r}')\Delta B(\vec{r}')\psinc(\vec{r}')\nonumber \\
          &+ \I\omega\int_{\Sigma}\out^{\dagger}\frac{Q}{4}\int_{V}\Gamma_{\rm bg}(\vec{r},\vec{r}')\Delta B(\vec{r}')\RQNM(\vec{r}')\vec{a}\,.
	\label{eq:full_exp}
\end{align}
The first term simply isolates the outgoing components of the incident field; from \eqref{incs}, this term equals $S_{\rm bg} \cin$.

The second and third terms start with polarization-field terms inside the scatterer, $\Delta B \psi_{\rm inc}$ and $\Delta B \RQNM \vec{a}$, respectively, convolve with the background Green's function to yield radiated (outgoing) fields on the bounds surface $\Sigma$, and then computes the power-normalized overlap with the outgoing-wave basis functions, $\Phi_{\rm out}$. We can simplify these expressions via reciprocity. We assume that the background medium is a reciprocal material. (If it is not, then one can simply use generalized reciprocity~\cite{Kong1975} and slightly modified versions of the expressions below.) Reciprocity in the background means that the background Green's function has a symmetry under reversal of its position arguments, i.e., transposition of source and receiver locations. Typically that would be written $\Gamma_{\rm bg}^T(\vec{r},\vec{r}') = \Gamma_{\rm bg}(\vec{r}',\vec{r})$. However, this exact expression relies on either free-space or periodic boundary conditions, and does not apply, for example, in scattering scenarios with Bloch-periodic boundary conditions. With full generality, for \emph{any} boundary conditions, the reciprocal scenario represented by the transpose and argument reversal of the Green's function is given by \emph{time-reversing} the scattering channels, which we can encode in the Green's function as the relation
\begin{equation}
    \Gamma_{\mathrm{bg}}^{T}(\vec{r},\vec{r}') = \Gamma^{\rm TR}_{\mathrm{bg}}(\vec{r}',\vec{r})\,,
    \label{eq:recip}
\end{equation}
where the ``TR'' superscript implies time-reversal of the channels. This relation does not require any time-reversal symmetry of the scatterer itself. This definition includes both the usual symmetry of the Green's function for free-space or periodic boundary conditions, as well as the special inner product defined in the Bloch-periodic case~\cite{lecamp2007theoretical}. 

We now simplify the third expression on the right-hand side of \eqref{full_exp}, which will imply a similar simplification for the second expression as well. First, we observe that adding in any constant multiple of $\in$ to $\out$ in the overlap integrals will not change the value of that integral, due to the orthogonality $\inner{\in}{\out} = 0$ and the scattered field being always outgoing. 
    As a result , we add $\frac{\alpha}{\beta} \in$ to $\out$, and this sum becomes $\frac{1}{\beta}\inc$ by \eqref{incbasis}, so we can replace $\out^\dagger$ with $\bb{\frac{1}{\beta}\inc}^\dagger$. Now we can work with a regular field $\inc$, instead of the singular field $\out$, which will enable a simplification below. The basis functions in $\inc^*$ are related to their time-reversed partners by $\inctr = P\inc^*$, where $P = \bb{\mat{1&0\\0&-1}}$. Using the fact that $P^2 = \mathbb{I}$, we can write $\bb{\frac{1}{\beta}\inc}^\dagger$ as $\frac{1}{\beta^*}(P \inctr )^T$. To use reciprocity in the Green's function part of the expression in \eqref{full_exp}, we want to take its transpose. Upon using the reciprocity relation, \eqref{recip}, the position arguments of the Green's function can be reversed, in which case one can interpret the expression in a new way: fields on the bounding surface $\Sigma$ are transported \emph{into} the scatterer, by the background Green's function, where they are overlapped with the fields inside the scatterer. Working through this intuition mathematically, we find:
\begin{align}
    &\cb{\int_{\Sigma}\frac{\I\omega}{\beta^*}(P \inctr )^T (\vec{r})\frac{Q}{4}\int_{V}\Gamma_{\rm bg}(\vec{r},\vec{r}')\Delta B(\vec{r}')\RQNM(\vec{r}')\vec{a}}^{T}\nonumber\\
        &=\frac{\I\omega\vec{a}^T}{4\beta^*}\int_{V}\RQNM^{T}(\vec{r}')\Delta B^{T}(\vec{r}') \int_{\Sigma}\Gamma^{\rm TR}_{\rm bg}(\vec{r}',\vec{r})Q P\inctr(\vec{r})\nonumber\\
        &=\frac{\I\omega\vec{a}^T}{4\beta^*}\int_{V}\RQNM^{T}(\vec{r}')\Delta B^{T}(\vec{r}') \inctr(\vec{r}') \nonumber\\
        &= \frac{\I\omega}{4\beta^*}(\inctr, \Delta B \RQNM) \vec{a}\,,
        \label{eq:deriv}
\end{align}
The initial expression is the third right-hand side expression of \eqref{full_exp}, with $\frac{1}{\beta^*}(P \inctr )^T$ replacing $\out^\dagger$. The first equality expression is the transpose of the initial with $\Gamma_{\rm bg}^{\rm TR}(\vec{r}',\vec{r})$ replacing $\Gamma_{\rm bg}^T(\vec{r},\vec{r}')$ via reciprocity, and $Q^T$ replacing $Q$ because it is complex symmetric. The second equality expression simplifies the integral on the right-hand side of the previous expression, $\int_{\Sigma}\Gamma^{\rm TR}_{\rm b}(\vec{r}',\vec{r})Q P\inctr(\vec{r}) = \inctr(\vec{r}')$, through the surface-equivalence principle~\cite{Kong1975,jin2011theory}. The fact that $\inctr$ is a traveling wave and free of singularities is crucial, otherwise the surface-equivalence principle cannot be applied. The term $Q P \inctr(\vec{r})$ are the equivalent surface currents that generate the incoming fields $\inctr$, and the convolution with the Green's function $\Gamma_{\rm bg}^{\rm TR}(\vec{r}',\vec{r})$ produces the fields $\inctr$ at points $\vec{r}'$ in the scattering body. The final equality expression is simply the transpose of the previous one, written in inner product notation, where we now clearly see that the lengthy expressions on the right-hand side of \eqref{full_exp} are proportional to simple unconjugated overlap integrals of the time-reversed incident fields with the fields $\Delta B \RQNM$, which are nonzero only in the scatterer. The specific nature of $\RQNM$ played no role in the derivation of \eqref{deriv}, and thus the simplification of the second term in \eqref{full_exp} is of exactly the same form but with the replacement $\RQNM \vec{a} \rightarrow \psinc$.

Having simplified each of the three terms in \eqref{full_exp}, we now have the second QCMT equation:
\begin{align}
\cout =& \cb{S_{\mathrm{bg}} + \frac{1}{4\alpha\beta^*}\I\omega (\inctr,\Delta B \inc)}\cin \nonumber\\
      &+\frac{1}{4\beta^*}\I\omega(\inctr,\Delta B \RQNM )\vec{a}\,.
	\label{eq:cqmt_second}
\end{align}
\Eqref{cqmt_second}, to our knowledge, has not been derived before. (See {\SM}  for the derivation without equivalence principle for structures  in vacuum.)
Intuitively, the first and third terms represent the direct background process from incoming to outgoing waves, and the radiation from QNMs to outgoing waves, respectively. Interestingly, the second term represents a Born-like scattering term (it is the first term in a Born scattering series expansion) that apparently is not captured in the resonant response of the third term. \Eqref{cqmt_second} is the crucial QCMT equation that enables solution of the outgoing fields for a given input, and it will be the key to enabling an expression for the scattering matrix.

\section{Quasinormal coupled mode theory (QCMT)}
\label{sec:QCMT}
We can synthesize the two key results of the previous sections into our quasinormal coupled mode theory. To simplify comparisons with conventional CMT, we define frequency-dependent matrices $D(\omega)$ and $K(\omega)$ that play similar roles to $D$ and $K$ in conventional CMT:
\begin{align}
    D(\omega) &= \frac{\I\omega}{ \alpha}(\LQNM, \Delta B \inc)^T \label{eq:Kw} \\
    K(\omega) &= \frac{\I\omega}{4\beta^*}(\inctr,\Delta B \RQNM )\,. \label{eq:Dw}
\end{align}
With these matrices, we can write the key QCMT equations, \eqreftwo{cqmt_first}{cqmt_second}, as:
\begin{align}
    &\I N(\omega)(\Omega-\omega)\vec{a} = D^T(\omega)\cin \,, \label{eq:cqmt1} \\
    &\cout = \cb{S_{\mathrm{bg}} +\frac{\I\omega}{4\alpha\beta^*} (\inctr,\Delta B \inc)}\cin + K(\omega)\vec{a}\,. \label{eq:cqmt2}
\end{align}
Additionally, we can solve the first QCMT equation, \eqref{cqmt1}, for the quasinormal-mode amplitudes $\vec{a}$, insert the result into the second QCMT equation, \eqref{cqmt2}, and extract the QCMT scattering matrix:
\begin{align}
    S = S_{\mathrm{bg}} &+ \frac{\I\omega}{4\alpha\beta^*} (\inctr,\Delta B \inc) \nonumber \\
    &- \I K(\omega) \left[N(\omega)(\Omega-\omega)\right]^{-1} D^T(\omega)\,.
        \label{eq:cqmt_smat}
\end{align}
We see that \eqrefrange{cqmt1}{cqmt_smat} show a similar functional form to the analogous CMT equations, Eqs.~(\ref{eq:cmt1},\ref{eq:cmt2},\ref{eq:cmt_smat}). There are two key differences. First, the coupling matrices $K(\omega)$ and $D(\omega)$ are now frequency-dependent, with a possible additional frequency dependence arising in dispersive media from the matrix $N(\omega)$.  This frequency dependence is critical to accurate simulations, as we show in \secref{examples}; in the time domain, they indicate that the coupling operators are convolutions, as we discuss in \secref{cmt_acc}. The second key difference is the appearance of the second term on the right-hand sides of \eqref{cqmt2} and \eqref{cqmt_smat}, which is proportional to the overlap of the time-reversal-generated incident waves with the incident field, in the domain of the scatterer. This Born-scattering term arises only in the presence of a scatterer (i.e. $\Delta B \neq 0$ everywhere), and yet is part of the ``direct'' scattering process, a term that has no counterpart in conventional CMT.

One can similarly ask whether the QCMT equations satisfy conservation laws similar to those of \eqrefrange{KD}{Dbg} of conventional CMT. With reciprocal materials and outgoing channel functions that are time-reversed partners of the incoming channel functions, one can see from \eqreftwo{Kw}{Dw} that $D(\omega)$ and $K(\omega)$ will be identical,
\begin{align}
    D(\omega) = K(\omega) \qquad \textrm{(reciprocal)}
\end{align}
up to the numerical factor $\alpha/4\beta^*$. However, that is as far as one can go with simple QCMT conservation laws. The analog of \eqref{KdagK}, $K^\dagger K = 2 \Im \Omega$, does not hold. Intuitively, that equality is a statement that the mode-energy decay rate equals the power in the outgoing channels, and certainly one could codify such a statement in a Poynting-flux evaluation of the CMT quantities. However, $\Im \Omega$ does not determine the mode-energy decay rate at arbitrary frequency $\omega$ in the general scenario when the coupling matrices are frequency-dependent. Similarly, there is no simple analog of \eqref{Dbg}, $\Sbg D^* = -D$, which enforces a relation between the background scattering matrix and the coupling matrix $D$. A key impediment is the presence of the Born term in the second CMT equation, \eqref{cqmt2}, which augments the background with a scatterer- and frequency-dependent matrix. 

\section{Pole expansion representations}
\label{sec:pole}
The key equations derived to this point are \eqrefrange{cqmt1}{cqmt_smat}, which are the two QCMT equations and the corresponding scattering matrix, in order. A key distinction between conventional CMT and these equations is that the QCMT matrices are frequency-dependent, and require use of an overlap matrix $N(\omega)$ and its inverse. In this section we show how Mittag--Leffler pole expansions allow for removal of the $N$ matrix and simplification of the frequency-dependent matrices.

For our purposes, we can use the following form of a Mittag--Leffler expansion~\cite{whittaker2020course,ablowitz2003complex}: given a meromorphic function $f(z)$ with simple poles $z_1, z_2, \ldots, z_n,\ldots$, no pole at 0, and $\lim_{n\rightarrow \infty}z_n = \infty$, one can expand $f(z)$ around $z=0$ as
\begin{equation}
	f(z) = f(0) + h(z) + \sum_{n=1}^\infty\frac{\mathrm{Res}(f(z_n))}{z_n} + \sum_{n=1}^\infty\frac{\mathrm{Res}(f(z_n))}{z-z_n} \,,
	\label{eq:ML_exp}
\end{equation}
where $\mathrm{Res}(f(z_n))$ is the residue of $f(z)$ at $z_n$ and $h(z)$ is an entire function. By definition, $h(0) = 0$. Furthermore, if $f(z)$ is bounded as $z$ goes to complex infinity, then $h(z)$ is zero everywhere. In this section we emphasize only new expressions; in the SM we show that Mittag--Leffler expansions at various stages of the QCMT formulation unify many seemingly different expressions that have previously appeared across the literature.

\subsection{Simplified QCMT equation}
A first use of the Mittag--Leffler expansion is to simplify the first QCMT equation, \eqref{cqmt1}. We can start with the volume-integral expression relating the scattered field at a point $\vec{r}$ in the scatterer to the incident field at a point $\vec{r}'$ in the scatterer:
\begin{equation}
	\pss(\vec{r}) = \I\omega\int \Gamma(\vec{r},\vec{r}',\omega)\Delta B(\vec{r}',\omega)\psinc(\vec{r}')\,.
	\label{eq:Ev_eqv}
\end{equation}
To identify all possible poles in the response, we require knowledge of the frequency dependence of the permittivity, which by the Kramers--Kronig relations~\cite{nussenzveig1972causality} (or a pole-expansion representation~\cite{luo2013surface,raman2013upper}), can be written
\begin{align}
    B(\omega) = B_{\infty} - \sum_n \frac{\sigma_n}{\omega-\omega_n},
    \label{eq:epsw}
\end{align}
where the $\omega_n$ are the complex-frequency material poles (to be distinguished from the quasinormal-mode frequencies $\titwi{i}$) and the $\sigma_n$ are matrix-valued residues. The Green's function can be decomposed into a summation over the quasinormal-mode resonances~\cite{muljarov2016exact,lalanne2018light},
\begin{align}
    \Gamma(\vec{r},\vec{r}',\omega) = \sum_i \frac{\psi_{R,i}(\vec{r})\psi_{L,i}^T(\vec{r}')}{\I(\titwi{i} - \omega)}.
    \label{eq:GEE}
\end{align}
From \eqref{GEE} and a known sum rule~\cite{muljarov2016resonant}, one can show that $\Gamma$ evaluated at any material pole is 0, i.e. $\Gamma(\vec{r},\vec{r}',\omega_n) = 0$, still considering $\vec{r}$ and $\vec{r}'$ at points inside the scatterer. Physically this must be true because at any material pole the permittivity diverges and the materials acts as a perfect conductor, such that the field within the scatterer is 0. The term $\Gamma \Delta B$ in \eqref{Ev_eqv} contains the frequency-dependent term $\Gamma \sigma_n / (\omega - \omega_n)$; applying ML to that term yields:
\begin{eqnarray}
        &&\Gamma(\vec{r},\vec{r}',\omega) \sum_n \frac{\sigma_n}{\omega-\omega_n}	\nonumber\\
        &=&  \sum_{i,n} \frac{\psi_{R,i}(\vec{r})\psi_{L,i}^T(\vec{r}')}{\I(\titwi{i} - \omega)}\frac{\sigma_n}{(\titwi{i} -  \omega_n)} \nonumber\\
        &&- \sum_{i,n} \frac{1}{\I}\psi_{R,i}(\vec{r})\psi_{L,i}^T(\vec{r}')\frac{\sigma_n}{\titwi{i}(\titwi{i} -  \omega_n)}
\nonumber\\
&& + \Gamma(\vec{r},\vec{r}',\omega_n)\sum_n\frac{\sigma_n}{\omega -  \omega_n}-\Gamma(\vec{r},\vec{r}',0)\sum_n\frac{\sigma_n}{\omega_n}\nonumber\\
	&=&  \sum_{i,n} \frac{\psi_{R,i}(\vec{r})\psi_{L,i}^T(\vec{r}')}{\I(\titwi{i} - \omega)} \sum_n \frac{\sigma_n}{(\titwi{i} -  \omega_n)}\,, \nonumber \\
        &=& \sum_{i,n} \frac{\psi_{R,i}(\vec{r})\psi_{L,i}^T(\vec{r}')}{\I(\titwi{i}-\omega)}( B_{\infty} - B(\titwi{i}))
        \label{eq:new_exp}
\end{eqnarray}
where only the first term remains in the expansion because the fourth term contains $\Gamma(\vec{r},\vec{r}',0)$, the third term contains $\Gamma(\vec{r},\vec{r}',\omega_n)$, and the second term is proportional (via partial-fraction expansion, cf. {\SM}) to the difference $\Gamma(\vec{r},\vec{r}',0) - \Gamma(\vec{r},\vec{r}',\omega_n)$, all of which are zero due to the identities discussed above, $\Gamma(\vec{r},\vec{r}',0) = \Gamma(\vec{r},\vec{r}',\omega_n) = 0$. We define a basis matrix $\LQNMT$ with elements $\left[\Delta B(\titwi{i})\right]^T \psi_{L,i}$. Inserting this expansion back into \eqref{Ev_eqv}, we find a simple expansion expression:
\begin{equation}
    (\Omega-\omega)\vec{a} = \frac{\omega}{\alpha}(\LQNMT, \inc)\cin \,,
    \label{eq:cqmt_firstv2}
\end{equation}
This expression simplifies the frequency dependencies of \eqref{cqmt_first}, as it does not contain the frequency-dependent overlap matrix $N(\omega)$, nor does it require re-evaluating the material constant $B$ at every frequency $\omega$, and it is very similar to the first CMT equation, albeit with a frequency-dependent prefactor. To our knowledge, \eqref{cqmt_firstv2} has not been derived before. Similarly, substituting $\Gamma$ in \eqref{GEE} into \eqref{Ev_eqv} produces a version of \eqref{cqmt_first} without $N(\omega)$, so we can similarly drop $N(\omega)$ in the expansions to follow. Furthermore, applying the same procedures to $\omega\Gamma \Delta B$, we obtain the expansion expression in Refs.\cite{muljarov2016resonant, yan2018rigorous} (cf. {\SM}). This method also makes the equivalence evident among other expansion expressions highlighted in \citeasnoun{lalanne2018light}.

\subsection{Resonant expansion of S matrices}
The S matrix expression in \eqref{cqmt_smat} is similar to a pole expansion via resonance frequencies, except for the presence of the $N$ matrix and the frequency dependencies of the $K$ and $D$ matrices. We can apply the Mittag--Leffler expansion to \eqref{cqmt_smat} to obtain a simpler form. As shown in the SM, the result utilizes frequency-independent matrices $\tit{K}$ and $\tit{D}$, defined by $\tit{K}_{ij} = K_{ij}(\titwi{j})$ and $\tit{D}_{ij} = D_{ij}(\titwi{i})$\,. Then \eqref{cqmt_smat} becomes
\begin{equation}
    S =  \equals{S_{\rm bg}(\omega = 0) +H(\omega) + \I\tit{K}\Omega^{-1}\tit{D}^T}{\mbox{``background part''}} - \equals{\I\tit{K}(\Omega - \omega)^{-1}\tit{D}^T}{\mbox{``resonant part''}}\,,
	\label{eq:s_mat_pole}
\end{equation}
where $H$ is a frequency dependent background term (generalizing $h$ from \eqref{ML_exp} to a matrix), whose elements are all entire functions of $\omega$, containing the Born scattering term and the difference between $\tit{K}(\Omega - \omega)^{-1}\tit{D}^T$ and $K(\Omega - \omega)^{-1}D^T$.
The term with the inverse of $(\Omega - \omega)$ will be the dominant contributor near resonances, while the remaining terms can be considered a non-resonant ``background.'' With the frequency dependencies of $K$ and $D$ removed (and $N$ removed altogether), the resonant term is now purely a complex Lorentzian form. For simplicity, we assumed the scattering poles do not exactly coincide with the material poles. The key hurdle to using \eqref{s_mat_pole} is the presence of the unknown function $H(\omega)$. Without knowledge of $H(\omega)$, one can only determine $H(\omega)$ from \eqref{cqmt_smat} (cf. {\SM}), rendering \eqref{s_mat_pole} redundant. For choices of channel basis functions such that $S(\omega)$ goes to zero at infinity everywhere in the complex plane (as is possible in the Fabry--Perot example in \secref{examples}), however, $H(\omega) = 0$ and \eqref{s_mat_pole} represents a dramatic simplification of the QCMT scattering matrix.



\begin{figure*}[htb]
    \includegraphics[width=0.8\linewidth]{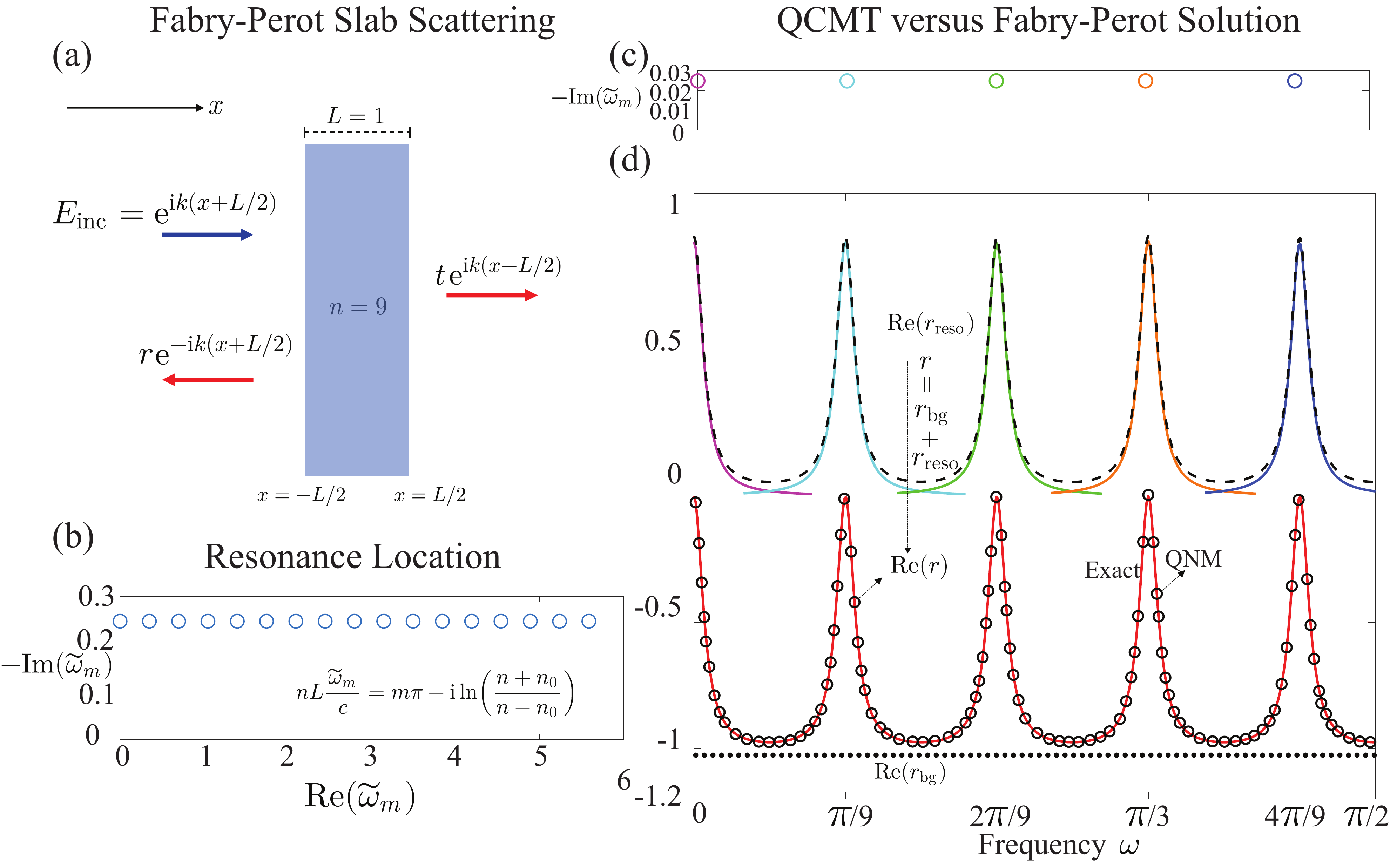}
    \caption{(a) A plane wave in vacuum is incident upon a Fabry-Perot slab of thickness $L=1$ (alternatively, all frequencies are scaled by $1/L$ for arbitrary $L$) and refractive index $n=9$. The transmission and reflection coefficients of the slab are $t$ and $r$, respectively. The thickness-dependent phases of the channel functions ensure $H(\omega) = 0$ in  \eqref{s_mat_pole}. (b) Resonances $\titwi{m}$ in the complex frequency plane, computed analytically (inset). (c,d) Five resonances (c) and the real parts of the background (black dotted), resonant (colored solid, black dashed, top), and total (black circles) reflection coefficients computed by QCMT, as well as the exact expression (red solid).}
    \label{fig:fp_slab}
\end{figure*}

\section{QCMT Computations}
\label{sec:examples}
In this section, we demonstrate the accuracy of the QCMT theory. We compute wave-scattering solutions by \eqref{cqmt_smat} from a Fabry-Perot slab as well as from a three-dimensional Mie sphere, two cases where one can easily compare against exact solutions. In each example, we demonstrate that it is critical to accurately model the contribution of the background term in \eqref{s_mat_pole} (including the Born scattering term), which was missing from previous quasinormal-mode descriptions of the scattering matrix~\cite{alpeggiani2017quasinormal,weiss2018calculate}. In the sphere case, we go a step further and compare the exact and QCMT results with the best possible CMT model of the scattering process. We show that CMT \emph{cannot} accurately model the scattering response, as even the best CMT approximation is highly inaccurate.

The Fabry--Perot example is depicted in \figref{fp_slab}(a). The basis functions for the incoming and outgoing waves are plane waves. The basis-function phases are defined relative to each interface, ensuring that the scattering matrix decays to zero at infinity everywhere in the complex plane, allowing us to use the simpler pole-expansion expressions of \secref{pole}. In \figref{fp_slab}(b) we plot the resonances of the slab (with refractive index $n=9$), which can be found analytically~\cite{lalanne2018light} by the inset expression. (The high value of $n$ makes the resonances more distinct; one can expect equivalent or even higher accuracy for smaller refractive indices.) We consider normal incidence where both $s$ and $p$ polarizations exhibit identical response. The reflection coefficients computed by QCMT are shown in \figref{fp_slab}(d), where each of the resonant contribution $r_{\rm reso}$, the background contribution $r_{\rm bg}$, and {the total reflection coefficient $r$ are shown. For the resonant-only reflection coefficient, the dashed black line shows the total computed $r_{\rm reso}$, while the solid lines depict the isolated contributions of each individual resonance. The background reflection coefficient, which our method predicts exactly, provides an important shift in the reflection coefficient that is critical to getting the correct final answer. The total reflection coefficient is depicted by the open circles, agreeing very well with the exact solution (red solid line). See {\SM} for further details of the channel function definitions and quasinormal modes. An accurate CMT model of the form of \eqrefrange{cmt1}{cmt_smat} for this example is impossible. Since the direct process here is perfect transmission, no CMT model can produce a reflection coefficient $r$ as in \figref{fp_slab}(d), with a few dips in a background of unity (total reflection).

\begin{figure*}[htb]
    \includegraphics[width=1.05\linewidth]{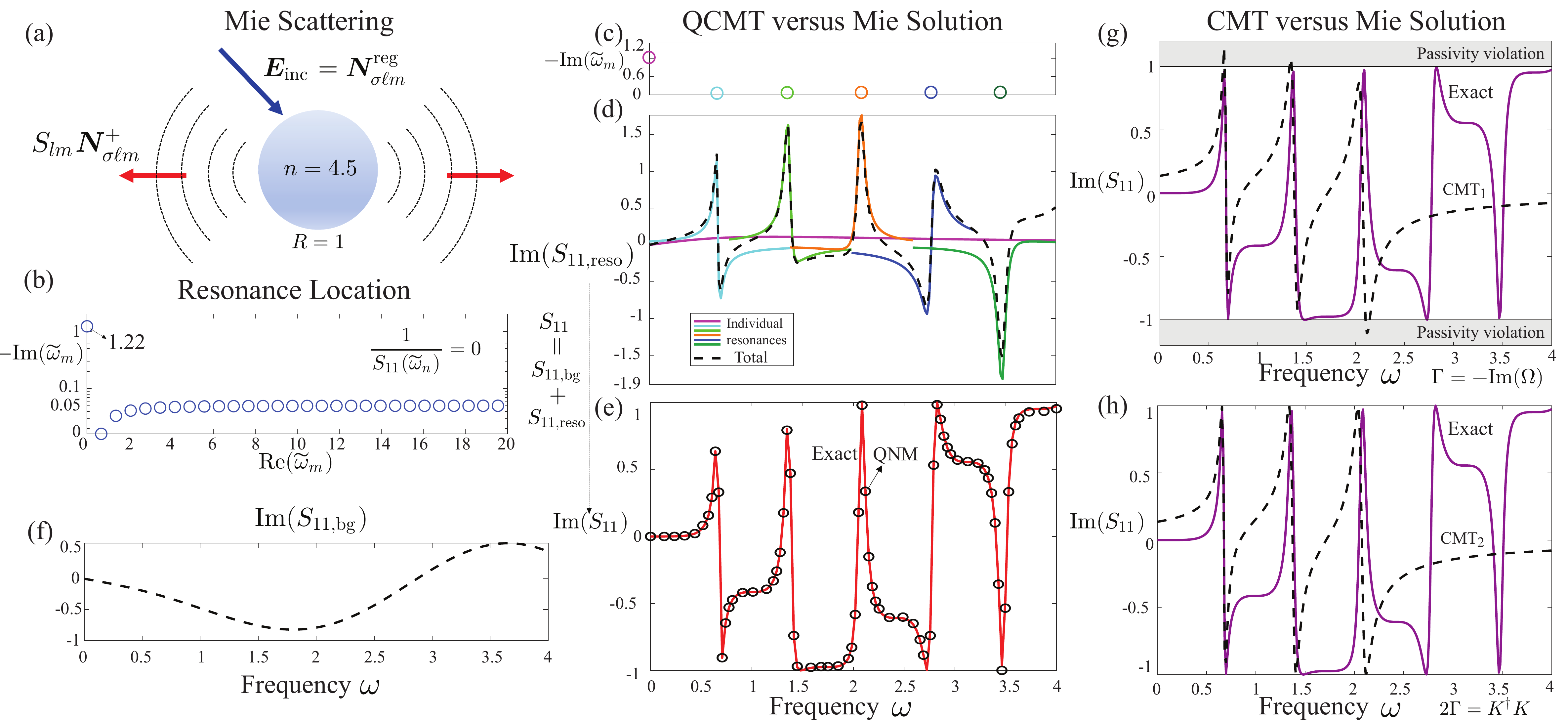}
    \caption{(a) An incoming spherical wave in vacuum scatters from a sphere of refractive index $n=4.5$ and radius $R=1$. (b) Computed resonances of the Mie sphere. (c) Six highlighted resonances. (d) Resonant contributions to the real part of the $S_{11}$ scattering-matrix coefficient from each resonance (solid colors), as well as in total (black dashed). (e) Total scattering-matrix coefficient by QCMT (black circles) and the exact solution (red solid lines). (f) The frequency-dependent background contribution. (g,h) CMT models of the Mie-scattering process, selecting the matrix $\Gamma$ equal to either $-\Im \Omega$ (g, CMT$_1$), in which case passivity can be violated in the CMT model, or $K^\dagger K / 2$ (h, CMT$_2$). Neither CMT model can accurately capture the response, even in the vicinity of the resonant peaks.}
    \label{fig:mie_sphere}
\end{figure*}

\Figref{mie_sphere} depicts the scattering by a sphere with refractive index $n=4.5$. Parts (a--e) of the figure are analogous to their counterparts in \figref{fp_slab}, and again show very high accuracy in the QCMT computations. The channel basis functions are vector spherical waves $\vect{N}_{\sigma\ell m}$ (\citeasnoun{kristensson2016scattering}). The indices $\ell$ and $m$ are the usual angular-momentum indices, while $\sigma$ separates the angular components into even and odd constituents. For a sphere the different angular momentum channels decouple, and here we consider the $\ell=1$ channel for the incident wave and scattering-matrix calculations. Unlike the Fabry--Perot case, the resonance locations cannot be found analytically. Instead, they are found as the poles of the sub-block of the $S$-matrix corresponding to the $\ell=1$ channel, $S_{11}$, i.e., the solution of the equation $\left(S_{11}(\titwi{m})\right)^{-1} = 0$. However, their QNM fields \emph{can} be obtained analytically using an appropriate rotation of the outgoing fields in the complex plane (cf. {\SM}). In \figref{mie_sphere}(a) we compare the imaginary part of $S_{11}$ constructed from \eqref{cqmt_smat} with the exact value in \figref{mie_sphere}(e). Unlike the Fabry-Perot case, there does not appear to be a choice of basis functions that makes the background part frequency-independent, and \figref{mie_sphere}(f) shows the oscillatory background term $S_{11,\textrm{bg}}$. Furthermore, we try to construct a CMT model of the form of \eqrefrange{cmt1}{cmt_smat} to the highest possible accuracy. In this CMT model, a single channel is coupled to multiple resonances represented by $\Omega$ as in \eqref{rqnm_eig}, so $K$ and $D$ are row vectors, with the number of elements set by the number of coupled resonances. Since the system here is reciprocal and lossless with $S_{\rm bg} =1 $, \eqreftwo{KD}{Dbg} must hold in the CMT model, requiring $K=D$ with purely imaginary elements. This fixes the phase of elements in $K$. Hence, we choose $K_m = \I\sqrt{-2\Im{\titwi{m}}}$ to accurately model the decay of the resonances. The choices of $\Gamma$ and $\Omega$ are not as straightforward as $K$ and $D$ because we have more modes than channels. This means $\Gamma = -\mathrm{Im}(\Omega)$ and $\Gamma = \frac{1}{2} K^\dagger K$ cannot be satisfied simultaneously. ($K^\dagger K $ is rank 1 with the single incoming/outgoing channel.) In \figref{mie_sphere}(g), with model ``CMT$_1$,'' we choose $\Gamma = -\mathrm{Im}(\Omega)$, in which case $K^\dagger K \ne 2\Gamma$ and energy conservation is violated. In \figref{mie_sphere}(h), for model ``CMT$_2$,'' we choose $\Gamma = \frac{1}{2} K^\dagger K$, in which case $\Gamma \ne -\mathrm{Im}(\Omega)$ (which is the choice in \citeasnoun{hsu2014theoretical}). In both cases, the agreement is poor between the CMT model and the exact soution, demonstrating the inability of the CMT model to accurately capture the multi-resonant scattering response.

\section{When is CMT accurate?}
\label{sec:cmt_acc}
The results of the previous section prompt a more general question about the validity of conventional CMT. Our exact QCMT theory allows us to uniquely answer this question. The QCMT theory simplifies to conventional CMT when the following conditions hold: (1) the Born-scattering background term is small, and (2) the coupling strengths $D(\omega)$ and $K(\omega)$ must be approximately frequency-independent over the bandwidth of interest. The second condition is a more precise statement of the well-understood requirement~\cite{haus1984waves,joannopoulosphotonic} that CMT requires high-quality-factor, well-separated resonances. 

The time-domain versions of the CMT equations of \eqreftwo{cmt1}{cmt2} are~\cite{suh2004temporal}
\begin{align}
    \frac{\D }{\D t}\vec{a}(t)&= -\I\Omega\vec{a}(t)  + D^{T}\cin(t) \label{eq:tcmt1} \\
    \cout(t) &= S_{\rm bg}\cin(t)  + K\vec{a}(t) \label{eq:tcmt2}\,,
\end{align}
which can be interpreted as the inverse Fourier transform of the frequency-domain equations, \eqreftwo{cmt1}{cmt2}. To identify the approximations inherent to the CMT equations, we can find the inverse Fourier transform of the frequency-domain QCMT equations, \eqreftwo{cqmt1}{cqmt2}; for nondispersive media (for which $N(\omega)$ is the identity), analogous manipulations yield
\begin{align}
    \frac{\D}{\D t}\vec{a}(t) =& -\I\Omega \vec{a}(t) +   \int\D t' D^T(t-t')\cin(t') \,,
    \label{eq:cqmt_first_time} \\
    \cout(t) =& \int\D t' \left\{S_{\rm bg}(t-t') + E(t-t')\right\}\cin(t') \nonumber \\
             &\quad\quad +  \int\D t' K(t-t')\vec{a}(t')\,,
             \label{eq:cqmt_second_time}
\end{align}
where $E(t)$ denotes the inverse Fourier transform of the Born scattering term. There are two prominent differences between the QCMT time-domain equations, \eqreftwo{cqmt_first_time}{cqmt_second_time}, and the CMT time-domain equations, \eqreftwo{tcmt1}{tcmt2}. First, the Born-scattering $E(t)$ term is a background contribution that is not accounted for in conventional CMT. (It cannot simply be lumped into $S_{\rm bg}$, which by definition is defined in the absence of any scatterer, and thus is scatterer-independent.) Second, one can see that all of the linear relations between the mode amplitudes and the incoming- and outgoing-wave coefficients are convolutions in time, which is required due to the frequency-dependence of the relevant matrices in the frequency domain.

Consequently, one can conclude that CMT will be valid when two conditions are met: when the Born term $E$ is negligible (i.e., single-pass scattering excites a much smaller amplitude than the incident wave), and when the coupling matrices such as $D(t)$ and $K(t)$ are sharply peaked in time; mathematically, one recovers the conventional CMT equations by neglecting $E(t)$ and assuming all other matrices are delta functions in time, or dispersionless. 

In the frequency domain we can more precisely identify the condition in which the coupling matrices can be treated as constants. Consider for example the matrix $D(\omega)$, whose $i^{\rm th}$ row $D_i(\omega)$ corresponds to the coupling strengths between resonance $i$, with resonant frequency $\titwi{i} = \omega_i - \I\Gamma_i$, and all incoming-wave amplitudes. This row can be approximated as a constant row $D_i(\titwi{i})$ if the next term in the Taylor expansion is sufficiently small. That term will be proportional to $\omega - \titwi{i}$ (and the first derivative of $D_i$); for a mode with high quality factor $Q$, the frequencies of interest will occupy a bandwidth proportional to $1/Q$ that will reduce the size of $|\omega - \titwi{i}|$. For high-quality-factor, well-separated modes, conventional CMT will apply. Beyond this limit, QCMT is required.



\section{Conclusions}
We have developed a QCMT framework as an exact generalization of CMT. Compared with CMT, the coupling constants between resonances and channels are frequency dependent and an extra non-resonant term, previously missing in CMT, appears. This framework reveals the underlying structure of scattering matrices and enables the exact decomposition and analysis of the response due to individual resonances in a  complex scattering problem. It also provides guidelines for the usage of CMT, allowing a systematic approximation from the exact theory. Looking forward, this work opens multiple avenues. QCMT can serve as a modeling paradigm for the design of complex nanophotonic structures, for applications ranging from metasurfaces~\cite{yu2014flat,aieta2015multiwavelength,lalanne2017metalenses,khorasaninejad2017metalenses,chung2020tunable,chung2020high,banerji2019imaging} to random media~\cite{hsu2017correlation,rotter2017light} to energy harvesting devices~\cite{de2012conversion,lenert2014nanophotonic}. There is an emerging interest in identifying fundamental limits to response in such structures~\cite{zhang2019scattering,presutti2020focusing}, and the QCMT framework could be ideal for identifying new bounds via the convenient mode/channel structure of the underlying equations. The QCMT framework could be paired with known sum rules on modal densities~\cite{barnett1996sum,shim2019fundamental}, or used in tandem with energy-conservation constraints~ \cite{molesky2020global,kuang2020computational,gustafsson2020upper} to identify the extreme limits of what is possible.

\bibliography{qcmt}

\clearpage
\ifarXiv
    

\section*{Supplementary material}

# Supplementary Materials: Quasinormal Coupled Mode Theory

Hanwen Zhang[1] and Owen D. Miller[1]

[1]*Department of Applied Physics, Yale University, New Haven, Connecticut 06511, USA*
(Dated: October 20, 2020)

## CONTENTS

## I. EXAMPLES OF S MATRIX QUASINORMAL MODE EXPANSION

In this section we provide details for the examples provided in the main text. The first example in Sec. I A is a Fabry-Perot slab, where the scattering channels are plane waves. The second example in Sec. I B is a Mie sphere, where the scattering channels are vector spherical waves. We list the explicit form of the channel functions and resonance mode expressions, through which the $S$ matrix is constructed. In both examples, the material susceptibility is nonmagnetic, so that $\Delta B$ only has nonzero $\Delta\epsilon$, and we can work primarily with the electric field $\boldsymbol{E}$ only. The magnetic field $\boldsymbol{H}$, if not shown, can be found by Maxwell equations directly. We work in dimensionless unit and set $c = 1$, so $\widetilde{\omega}_m = \widetilde{k}_m$.

### A. QNM expansion of Fabry-Perot slab

#### 1. QNM basis

For a Fabry-Perot slab configuration in Fig. 2(a), the normalized QNMs inside the slab of refractive index $n$ are given by ($s$ and $p$ polarizations are degenerate) [1]

$$\boldsymbol{E}_{\text{qnm},m} = \begin{cases} \frac{1}{n\sqrt{L}} \cos(n\widetilde{k}_m x)\hat{\boldsymbol{z}}, & \text{for } -L/2 < x < L/2, m \text{ even}, \\ \frac{1}{n\sqrt{L}} \sin(n\widetilde{k}_m x)\hat{\boldsymbol{z}}, & \text{for } -L/2 < x < L/2, m \text{ odd}, \end{cases} \quad (1)$$

with resonant frequencies

$$\widetilde{k}_m = \frac{1}{nL}\left[m\pi - i\ln\left(\frac{n+n_0}{n-n_0}\right)\right]. \quad (2)$$

For this one-dimensional example, the normalization integral is 0 outside of the slab, so a PML is not needed to normalize the QNMs.

### 2. Channel basis functions

The plane wave basis function has not been treated as systematically as the vector spherical waves and is normally used in a intuitive manner. As a result, first we need to formalize the plane wave basis to be consistent with the scattering framework developed in the main text. Here, we take the plane waves as incident basis $\Phi_{\text{inc}}$, which is a traveling wave regular throughout all space. We use Heaviside step function $\eta(x)$ with plane waves to represent the incoming basis $\Phi_{\text{in}}$ and outgoing basis $\Phi_{\text{out}}$, so that there are net power flow into/out the bounding surface $\Sigma$, which is normally two flat surfaces enclosing a unit cell of a periodic structure. The discontinuity due to $\eta(x)$ can be understood as the presence of additional sinks/sources, which is similar to the singularities in incoming and outgoing basis in vector spherical waves. Although here we only treat the plane wave basis without higher specular orders, the generalization is straightforward.

We choose $\boldsymbol{E}_{\text{in},1} = \eta(-x)\sqrt{2}e^{ik(x+L/2)}\hat{\boldsymbol{z}}$, $\boldsymbol{E}_{\text{in},2} = \eta(x)\sqrt{2}e^{-ik(x-L/2)}\hat{\boldsymbol{z}}$ as the incoming channel basis, and $\boldsymbol{E}_{\text{out},1} = \eta(x)\sqrt{2}e^{ik(x-L/2)}\hat{\boldsymbol{z}}$, $\boldsymbol{E}_{\text{out},2} = \eta(-x)\sqrt{2}e^{-ik(x+L/2)}\hat{\boldsymbol{z}}$, as the outgoing ones. One can easily check they are power orthonormal. Then we have

$$\begin{aligned}\boldsymbol{E}_{\text{inc},1} &= \boldsymbol{E}_{\text{in},1} + e^{ikL}\boldsymbol{E}_{\text{out},1} = \sqrt{2}e^{ik(x+L/2)}\hat{\boldsymbol{z}},\\ \boldsymbol{E}_{\text{inc},2} &= \boldsymbol{E}_{\text{in},2} + e^{ikL}\boldsymbol{E}_{\text{out},2} = \sqrt{2}e^{-ik(x-L/2)}\hat{\boldsymbol{z}},\end{aligned} \quad (3)$$

where $\eta(x)$ and $\eta(-x)$ adds up to unity and the discontinuity disappears. We can see that $\alpha = 1$ and $\beta = e^{ikL}$, due to the special phase choice.

### 3. S matrix construction

In the channel basis chosen above, the scattering matrix is of the form $S = \begin{pmatrix} t_1 & r_2 \\ r_1 & t_2 \end{pmatrix}$, and $S_{\text{bg}} = \begin{pmatrix} e^{ikL} & 0 \\ 0 & e^{ikL} \end{pmatrix}$ without the presence of the slab. Here the reflection coefficients $r_1 = r_2 = r$ and the transmission coefficients $t_1 = t_2 = t$, and the exact expressions are [2]

$$r = \frac{r_0(e^{2inkL}-1)}{1-r_0^2 e^{2inkL}}, \qquad t = \frac{t_0^2 e^{inkL}}{n(1-r_0^2 e^{2inkL})}, \quad (4)$$

where $r_0 = \frac{n-n_0}{n+n_0}$ and $t_0 = \frac{2n}{n+n_0}$.

The task, then, is to test whether the QNM expressions for the scattering matrix produce results that are consistent with the exact expressions of Eq. (4). In the QCMT framework, the $S$ matrix (as given in the main text), can be written

$$S = S_{\text{bg}} + \frac{i\omega}{4\alpha\beta^*}(\Phi_{\text{inc}}^{\text{TR}}, \Delta B \Phi_{\text{inc}}) - iK(\omega)\left[N(\omega)(\Omega - \omega)\right]^{-1}D^T(\omega). \quad (5)$$

The reflection and transmission coefficients are the components of the $S$-matrix: $r_1 = S_{11}$, $r_2 = S_{22}$, $t_1 = S_{21}$, and $t_2 = S_{12}$. Here the material is non-dispersive so $N(\omega) = \mathbb{I}$, the identity matrix. From the definitions of the basis functions and coupling matrices, and since given our channel definitions we have $\frac{1}{\alpha\beta^*} = e^{ikL}$, $\frac{1}{\alpha\beta^*}\boldsymbol{E}_{\text{inc},1}^{\text{TR}} = \boldsymbol{E}_{\text{inc},2}$ and

$\frac{1}{\alpha\beta^*}\boldsymbol{E}_{\text{inc},2}^{\text{TR}} = \boldsymbol{E}_{\text{inc},1}$, the QCMT reflection and transmission coefficients are given by

$$r_1 = \frac{1}{4}\mathrm{i}\omega(\boldsymbol{E}_{\text{inc},1}, \Delta\epsilon\boldsymbol{E}_{\text{inc},1}) + \frac{1}{4}(\mathrm{i}\omega)^2 \sum_m (\boldsymbol{E}_{\text{inc},1}, \Delta\epsilon\boldsymbol{E}_{\text{qnm},m}) \frac{1}{\mathrm{i}(\widetilde{\omega}_m - \omega)}(\boldsymbol{E}_{\text{qnm},m}, \Delta\epsilon\boldsymbol{E}_{\text{inc},1}), \tag{6}$$

$$r_2 = \frac{1}{4}\mathrm{i}\omega(\boldsymbol{E}_{\text{inc},2}, \Delta\epsilon\boldsymbol{E}_{\text{inc},2}) + \frac{1}{4}(\mathrm{i}\omega)^2 \sum_m (\boldsymbol{E}_{\text{inc},2}, \Delta\epsilon\boldsymbol{E}_{\text{qnm},m}) \frac{1}{\mathrm{i}(\widetilde{\omega}_m - \omega)}(\boldsymbol{E}_{\text{qnm},m}, \Delta\epsilon\boldsymbol{E}_{\text{inc},2}), \tag{7}$$

$$t_1 = \mathrm{e}^{\mathrm{i}kL} + \frac{1}{4}\mathrm{i}\omega(\boldsymbol{E}_{\text{inc},2}, \Delta\epsilon\boldsymbol{E}_{\text{inc},1}) + \frac{1}{4}(\mathrm{i}\omega)^2 \sum_m (\boldsymbol{E}_{\text{inc},2}, \Delta\epsilon\boldsymbol{E}_{\text{qnm},m}) \frac{1}{\mathrm{i}(\widetilde{\omega}_m - \omega)}(\boldsymbol{E}_{\text{qnm},m}, \Delta\epsilon\boldsymbol{E}_{\text{inc},1}), \tag{8}$$

$$t_2 = \mathrm{e}^{\mathrm{i}kL} + \frac{1}{4}\mathrm{i}\omega(\boldsymbol{E}_{\text{inc},1}, \Delta\epsilon\boldsymbol{E}_{\text{inc},2}) + \frac{1}{4}(\mathrm{i}\omega)^2 \sum_m (\boldsymbol{E}_{\text{inc},1}, \Delta\epsilon\boldsymbol{E}_{\text{qnm},m}) \frac{1}{\mathrm{i}(\widetilde{\omega}_m - \omega)}(\boldsymbol{E}_{\text{qnm},m}, \Delta\epsilon\boldsymbol{E}_{\text{inc},2}), \tag{9}$$

where $(A, B)$ here is $\int \mathrm{d}x A^T B$, a one-dimensional integral. It is obvious that $t_1 = t_2$ due to reciprocity. The symmetry $r_1 = r_2$ can be seen by a change of variable $x \to -x$ and noting $\boldsymbol{E}_{\text{qnm},m}$ is either odd or even in $x$. The constructed quantities here agree precisely with the exact expressions of Eq. (4), as shown in Fig. 2(d) of the main text.

### 4. Pole expansion of S matrix

Both $r$ and $t$ are bounded as $k$ goes to complex infinity, so one can apply Mittag-Leffler to $r$ and $t$ to obtain frequency-independent "background" and "resonant" terms. The QCMT $S$-matrix in such a case, as described in the main text, is given by the expression

$$S = S_{\text{bg}}(\omega = 0) + \mathrm{i}\widetilde{K}\Omega^{-1}\widetilde{D}^T - \mathrm{i}\widetilde{K}(\Omega - \omega)^{-1}\widetilde{D}^T, \tag{10}$$

where the first two terms comprise the background, while the third term is the resonant term. The background and resonant transmission and reflection coefficients have been derived for the specific case of Fabry–Perot [3], giving:

$$r_{\text{bg}} = t_{\text{bg}} = \frac{2\mathrm{i}}{(1-n^2)L}\sum_m \frac{1}{\widetilde{k}_m},$$

and

$$r_{\text{reso}} = \frac{2\mathrm{i}}{(1-n^2)L}\sum_m \frac{1}{k - \widetilde{k}_m}, \qquad t_{\text{reso}} = \frac{2\mathrm{i}}{(1-n^2)L}\sum_m \frac{(-1)^{m+1}}{k - \widetilde{k}_m}.$$

As shown in Fig. 2(b,d) of the main text, the QCMT calculations from Eq. (43) are in exact agreement with these expressions, while generalizing to arbitrary scattering bodies.

## B. QNM expansion of Mie sphere

For a spherical scattering body, the QNM fields can be expressed in terms of vector spherical waves (VSWs). Here we follow the convention of Ref. [4] for VSWs and write down QNM fields and channel functions. The resonant frequencies cannot be found analytically and must be numerically computed.

### 1. Vector spherical waves (VSWs)

VSWs in this convention have three indices, $\ell, m, \sigma$, which we collectively denote by $n$. Besides the common angular momentum numbers $\ell$ and $m$, the $\sigma$ index here takes value $e, o$, representing "even" and "odd" cases. This is because conventionally the spherical harmonics $Y_{\ell m}$, which are part of the VSW functions, have a factor $\mathrm{e}^{\mathrm{i}m\phi}$. However, here we separate it into $\cos(m\phi)$ and $\sin(m\phi)$, labeled by $\sigma$, taking value $e, o$ respectively. This makes the angular part purely real, a convenient choice under the unconjugated inner product for QNMs. Due to this extra index $\sigma$, we denote the spherical harmonics by $Y_n$ thereafter. They are given as

$$Y_n(\hat{\boldsymbol{r}}) = Y_{\sigma\ell m}(\hat{\boldsymbol{r}}) = \sqrt{\frac{\epsilon_m}{2\pi}}\sqrt{\frac{2\ell+1}{2}\frac{(\ell-m)!}{(\ell+m)!}} P_\ell^m(\cos\theta) \begin{Bmatrix} \cos\phi \\ \sin\phi \end{Bmatrix},$$

where $\hat{\boldsymbol{r}}$ is the unit position vector and $P_\ell^m(x)$ here are the associated Legendre polynomials, with

$$\sigma = e, o, \qquad m = 0, 1, 2, \ldots, \ell \qquad l = 0, 1, 2, \ldots, \qquad \epsilon_m = \begin{cases} 1, & m = 0 \\ 2, & m > 1 \end{cases}.$$

The angular part for a given $n$ consists of

$$\boldsymbol{A}_{1n} = \frac{1}{\sqrt{\ell(\ell+1)}} \nabla \times (\boldsymbol{r} Y_n(\hat{\boldsymbol{r}})), \tag{11}$$

$$\boldsymbol{A}_{2n} = \frac{1}{\sqrt{\ell(\ell+1)}} \boldsymbol{r} \nabla Y_n(\hat{\boldsymbol{r}}), \tag{12}$$

$$\boldsymbol{A}_{3n} = \hat{\boldsymbol{r}} \nabla Y_n(\hat{\boldsymbol{r}}), \tag{13}$$

which are an orthonormal set on the unit sphere since

$$\int_\Omega \boldsymbol{A}_{\tau n} \cdot \boldsymbol{A}_{\tau' n'} = \delta_{\tau\tau'} \delta_{nn'} \tag{14}$$

for $\tau, \tau' = 1, 2, 3$. We emphasize here again $\boldsymbol{A}_{\tau n}$ are real, so the above orthonormality relation is suitable for the unconjugated inner product.

The radial part consists of spherical Bessel functions $j_\ell(x)$, which is regular at $x = 0$, and spherical hankel functions of the first kind $h_\ell^{(1)}(x)$, which has an asymptotic form, for large $x$, proportional to outgoing spherical waves. As a result, this can be used to define outgoing channel functions, or to satisfy the radiation boundary condition of QNMs. Hence, we introduce the regular VSWs

$$\boldsymbol{v}_{1n}(k\boldsymbol{r}) = j_\ell(kr) \boldsymbol{A}_{1n}(\hat{\boldsymbol{r}}) \tag{15}$$

$$\boldsymbol{v}_{2n}(k\boldsymbol{r}) = \frac{(kr j_\ell(kr))'}{kr} \boldsymbol{A}_{2n}(\hat{\boldsymbol{r}}) + \sqrt{\ell(\ell+1)} \frac{j_\ell(kr)}{kr} \boldsymbol{A}_{3n}(\hat{\boldsymbol{r}}), \tag{16}$$

and the outgoing VSWs

$$\boldsymbol{u}_{1n}(k\boldsymbol{r}) = h_\ell^{(1)}(kr) \boldsymbol{A}_{1n}(\hat{\boldsymbol{r}}) \tag{17}$$

$$\boldsymbol{u}_{2n}(k\boldsymbol{r}) = \frac{(kr h_\ell^{(1)}(kr))'}{kr} \boldsymbol{A}_{2n}(\hat{\boldsymbol{r}}) + \sqrt{\ell(\ell+1)} \frac{h_\ell^{(1)}(kr)}{kr} \boldsymbol{A}_{3n}(\hat{\boldsymbol{r}}), \tag{18}$$

where $'$ represents the derivative of argument $kr$. With $h_\ell^{(1)}$ above replaced by $h_\ell^{(2)}(x) = (h_\ell^{(1)}(x))^*$, the spherical hankel functions of the second kind, the outgoing VSWs $\boldsymbol{u}_{\tau n}$ becomes the incoming ones $\boldsymbol{u}_{\tau n}^*$, which can be used to define incoming channel functions. Note that $j_\ell(x) = \frac{1}{2} h_\ell^{(1)}(x) + \frac{1}{2} h_\ell^{(2)}(x)$, so $\boldsymbol{v}_{1n}$ is a combination of incoming and outgoing fields, $\boldsymbol{v}_{\tau n} = \frac{1}{2} \boldsymbol{u}_{\tau n} + \frac{1}{2} \boldsymbol{u}_{\tau n}^*$ with $\alpha = \beta = \frac{1}{2}$.

### 2. Power normalization

For $\boldsymbol{E} = \boldsymbol{u}_{1n}$, the power flow out of the unit sphere is $\frac{1}{2} \operatorname{Re} \int_\Omega \boldsymbol{E} \times \boldsymbol{H}^* \cdot \mathrm{d}\boldsymbol{S} = \frac{1}{2k^2}$, which is not power normalized. Hence, normalization is needed when the channel functions are chosen. Here we define

$$\boldsymbol{N}_{\sigma\ell m}^{\mathrm{reg}} = \sqrt{2} k \boldsymbol{v}_{1n}, \quad \boldsymbol{N}_{\sigma\ell m}^+ = \sqrt{2} k \boldsymbol{u}_{1n}, \tag{19}$$

which is the notation used in Fig. 3(a) in the main text.

### 3. QNM and channel basis functions

With these functions define above, we are ready to write down QNMs and channel functions for the Mie spheres. We derive everything here for the $e$-polarization case, and the $h$-polarization counterpart will be similar.

We choose $\boldsymbol{E}_{\mathrm{out},n} = \sqrt{2} k \boldsymbol{u}_{1n}$, and $\boldsymbol{E}_{\mathrm{inc},n} = \sqrt{2} k \boldsymbol{v}_{1n}$, so that the outgoing channel basis is power normalized. If we denote $k_1 = nk$, the scattering matrix element can be found as [4]

$$S_{\ell m} = 1 + 2 T_\ell, \quad T_\ell = -\frac{j_\ell(kR)(k_1 R j_\ell(k_1 R))' - (kR j_\ell(kR))' j_\ell(k_1 R)}{h_\ell^{(1)}(kR)(k_1 R j_\ell(k_1 R))' - (kR h_\ell^{(1)}(kR))' j_\ell(k_1 R)}, \tag{20}$$

where $T_\ell$ is the transition matrix elements [5], which are more commonly used for Mie scattering.

The resonance frequencies $\widetilde{\omega}_m$, for a particular $n$, can be found by searching for zeros of $\frac{1}{S_{\ell m}}$ (or $\frac{1}{T_\ell}$). Next, we start to solve for QNM field expressions. Since $\boldsymbol{v}_{1n}$ is regular at the origin, and $\boldsymbol{u}_{1n}$ satisfy the radiation boundary condition,

$$\boldsymbol{E}_{\text{qnm},m} = \frac{1}{\sqrt{N_m}} \begin{cases} C\boldsymbol{v}_{1n}(n\widetilde{k}_m\boldsymbol{r}), & 0 \leq r < R, \\ \boldsymbol{u}_{1n}(\widetilde{k}_m\boldsymbol{r}), & R \leq r < \infty, \end{cases} \tag{21}$$

where $N_m$ is the normalization constant and the constant $C$ can be fixed by the continuity of electric field across $r = R$, which gives $C = \frac{h_\ell^{(1)}(kR)}{j_\ell(nkR)}$. By the relation that $\nabla \times \boldsymbol{v}_{1n}(k\boldsymbol{r}) = k\boldsymbol{v}_{2n}(k\boldsymbol{r})$, and $\nabla \times \boldsymbol{u}_{1n}(k\boldsymbol{r}) = k\boldsymbol{u}_{2n}(k\boldsymbol{r})$, the magnetic fields are

$$\boldsymbol{H}_{\text{qnm},m} = \frac{1}{\sqrt{N_m}} \begin{cases} -Cin\boldsymbol{v}_{2n}(n\widetilde{k}_m\boldsymbol{r}), & 0 \leq r < R, \\ -i\boldsymbol{u}_{2n}(\widetilde{k}_m\boldsymbol{r}), & R \leq r < \infty. \end{cases} \tag{22}$$

To work out the normalization constant $N_m$, one way to use the PML method as in the main text. This can be done by a coordinate stretching of $r \to \bar{r} = r + i\sigma(r)$, where $\sigma(r)$ turns on outside of the sphere [6]. Hence, by the orthonormality of $\boldsymbol{A}_{\tau n}$, the normalization integral $\int_V \epsilon \boldsymbol{E}_{\text{qnm},m}^2 - \mu \boldsymbol{H}_{\text{qnm},m}^2 = 2\int_V \epsilon \boldsymbol{E}_{\text{qnm},m}^2$ becomes a radial integral of $r$ only. The normalization condition gives

$$N_m = 2C^2 \int_0^R \mathrm{d}r\, r^2 n^2 j_\ell^2(n\widetilde{k}_m r) + 2\int_R^\infty \mathrm{d}\bar{r}\, \bar{r}^2 (h_\ell^{(1)}(\widetilde{k}_m \bar{r}))^2 \,. \tag{23}$$

For this particular example, this normalization method works, but it fails for purely imaginary $\widetilde{k}_m$. Besides, $\widetilde{k}_m \bar{r}$ has a large imaginary part for large $r$. The numerical evaluation of Bessel functions at large imaginary argument is unstable [7], and this could lead to potential numerical issues. Hence, an alternative normalization method can be found in Ref. [8], which is equivalent to the PML method in Ref. [9] and uses fields inside of the scatter only, can be adopted if numerical issues occur.

### 4. S matrix construction

Given the QNMs and channel functions as defined above, the QCMT $S$-matrix of Eq. (5) simplifies for the sphere to

$$S_{\ell m} = 1 + i\omega(\boldsymbol{E}_{\text{inc},n}, \Delta\epsilon \boldsymbol{E}_{\text{inc},n}) + (i\omega)^2 \sum_m (\boldsymbol{E}_{\text{inc},n}, \Delta\epsilon \boldsymbol{E}_{\text{qnm},m}) \frac{1}{i(\widetilde{\omega}_m - \omega)} (\boldsymbol{E}_{\text{qnm},m}, \Delta\epsilon \boldsymbol{E}_{\text{inc},n}) \,. \tag{24}$$

All inner product here, although 3-dimensional in nature, can be reduced into a radial integral by the orthonormality of $\boldsymbol{A}_{\tau n}$, as the normalization integral. This constructed $S_{\ell m}$ is plotted against the exact one of Eq. (20) in Fig. 3(e) in the main text.

## II. DERIVATION OF 2ND QCMT EQUATION IN SPECIAL CASES

In the main text, we derive the second QCMT equation,

$$\boldsymbol{c}_{\text{out}} = \left\{ S_{\text{bg}} + \frac{i\omega}{4\alpha\beta^*}(\Phi_{\text{inc}}^{\text{TR}}, \Delta B \Phi_{\text{inc}}) \right\} \boldsymbol{c}_{\text{in}} + K(\omega)\boldsymbol{a} \,, \tag{25}$$

by applying the equivalence principle to simplify the time-reversed channel functions. In this section, we provide an alternative proof, which is less general but perhaps more conventional, in the special cases of Fabry–Perot slabs and Mie spheres. The basic idea is to decompose the background Green's function according to the special symmetry of the scatterer.

## A. Fabry-Perot slabs

Consider a single slab of refractive index $n$ (as in Fig. 2(a) of the main text). We isolate the $x$ direction due to the $y, z$ transnational symmetry, and denote $(y, z)$ by $\boldsymbol{r}_\perp$, which is perpendicular to the $x$ direction. The total field $E$ satisfies the Helmholtz equation,

$$(\nabla^2 + k^2 \epsilon) E = 0, \qquad (26)$$

or

$$(\nabla^2 + k^2) E = -k^2 \Delta \epsilon E. \qquad (27)$$

Take the incoming field as $E_{\text{inc}} = \sqrt{2} e^{ik_x x}$ ($c_{\text{in},1} = 1$) and convert the above equation into the integral form and we have

$$E = E_{\text{inc}} + k^2 \int dS' dx' \frac{1}{4\pi} \frac{e^{ik|\boldsymbol{r} - \boldsymbol{r}'|}}{|\boldsymbol{r} - \boldsymbol{r}'|} \Delta \epsilon(x') E(x'), \qquad (28)$$

with $x$ integral isolated. As $\frac{1}{4\pi} \frac{e^{ik|\boldsymbol{r} - \boldsymbol{r}'|}}{|\boldsymbol{r} - \boldsymbol{r}'|}$ can be written as [10]

$$\frac{1}{4\pi} \frac{e^{ik|\boldsymbol{r} - \boldsymbol{r}'|}}{|\boldsymbol{r} - \boldsymbol{r}'|} = \int \frac{dk'_z}{2\pi} \frac{dk'_y}{2\pi} e^{i\boldsymbol{k}'_\perp \cdot (\boldsymbol{r}_\perp - \boldsymbol{r}'_\perp)} \frac{i}{2\sqrt{k^2 - k'^2_\perp}} e^{i\sqrt{k^2 - k'^2_\perp}|x - x'|}, \qquad (29)$$

the $S'$ integral produces a $\delta(\boldsymbol{k}'_\perp)$ and we have

$$E = \sqrt{2} e^{ikx} + \frac{ik}{2} \int_{-L/2}^{L/2} dx' e^{ik|x - x'|} \Delta \epsilon(x') E(x'). \qquad (30)$$

At $x > L/2$, we can write $E = t \sqrt{2} e^{ik_x x}$. We then have

$$t = 1 + \frac{ik}{4} \int dx' \sqrt{2} e^{-ikx'} \Delta \epsilon(x') E(x'). \qquad (31)$$

Similarly, at $x < -L/2$, we write $E(x) = \sqrt{2} e^{-ikx} + r \sqrt{2} e^{ikx}$. We then have

$$r = \frac{ik}{4} \int dx' \sqrt{2} e^{ikx'} \Delta \epsilon(x') E(x'). \qquad (32)$$

If we express $E = \sqrt{2} e^{ikx} + a_m \sum_m E_{\text{qnm},m}$, where $a_m$ is the element of the expansion coefficients $\boldsymbol{a}$, we obtain the second QCMT equation, Eq. (25), for $t$ and $r$, with outgoing channel functions $\eta(x) \sqrt{2} e^{ikx}$ and $\eta(-x) \sqrt{2} e^{-ikx}$. (See Section I A for the channel function definition.) If we solve for $\boldsymbol{a}$ from the first QCMT equation, $t$ and $r$ in Eq. (31) and Eq. (32) will be identical to Eq. (8) and Eq. (6), except for a phase factor of $e^{ikL}$.

## B. Mie spheres

Consider a single spherical scattering body (as in Fig. 3(a) of the main text). The total field $\boldsymbol{E}$ satisfies

$$\nabla \times \nabla \times \boldsymbol{E} - \epsilon k^2 \boldsymbol{E} = 0, \qquad (33)$$

or

$$\nabla \times \nabla \times \boldsymbol{E} - k^2 \boldsymbol{E} = \Delta \epsilon k^2 \boldsymbol{E}. \qquad (34)$$

For incident field $\boldsymbol{E}_{\text{inc}}$, we convert the equation into an integral equation

$$\boldsymbol{E} = \boldsymbol{E}_{\text{inc}} + k^2 \int G_{EE}(\boldsymbol{r}, \boldsymbol{r}') \Delta \epsilon \boldsymbol{E}(\boldsymbol{r}'), \qquad (35)$$

where $G_{EE} = (\mathbf{1} + \frac{1}{k^2}\nabla\nabla)\frac{\mathrm{e}^{\mathrm{i}k|\mathbf{r}-\mathbf{r}'|}}{4\pi|\mathbf{r}-\mathbf{r}'|} = \mathrm{i}k\sum_{n,\tau=1,2}\mathbf{u}_{\tau n}(k\mathbf{r}_>)\mathbf{v}_{\tau n}(k\mathbf{r}_<)$ [4]. Here $\Delta\epsilon$ has spherical symmetry, thus angle independent, so orthonormality of $\mathbf{A}_{\tau n}$ enable us to isolate each different $\tau$ and $n$ in $\mathbf{E}_{\mathrm{inc}}$. For $\mathbf{E}_{\mathrm{inc}} = \sqrt{2}k\mathbf{v}_{1n}$ ($c_{\mathrm{in},n} = c_{\mathrm{out},n} = \frac{1}{2}$, $c_{\mathrm{inc},n} = 1$) and $\mathbf{r}$ outside of the sphere, we have

$$\mathbf{E}(\mathbf{r}) = \sqrt{2}k\underbrace{\mathbf{v}_{1n}(\mathbf{r})}_{=\frac{1}{2}\mathbf{u}_{1n}+\frac{1}{2}\mathbf{u}_{1n}^*} + \mathbf{u}_{1n}(\mathbf{r})\mathrm{i}k^3\int \mathbf{v}_{1n}(\mathbf{r}')\cdot\Delta\epsilon\cdot\mathbf{E}(\mathbf{r}') \tag{36}$$

$$= \frac{1}{2}\sqrt{2}k\mathbf{u}_{1n}^*(\mathbf{r}) + \frac{1}{2}\sqrt{2}k\mathbf{u}_{1n}(\mathbf{r})\underbrace{\left(1 + \mathrm{i}k\int\sqrt{2}k\mathbf{v}_{1n}(\mathbf{r}')\cdot\Delta\epsilon\cdot\mathbf{E}(\mathbf{r}')\right)}_{=S_{\ell m}} \tag{37}$$

If we express $\mathbf{E} = \sqrt{2}k\mathbf{v}_{1n} + a_m\sum_m \mathbf{E}_{\mathrm{qnm},m}$, where $a_m$ is the element of the expansion coefficients $\mathbf{a}$, we obtain the second QCMT equation, Eq. (25), for $e$-polarized $S_{\ell m}$ for Mie spheres, with power normalized outgoing channel $\sqrt{2}k\mathbf{u}_{1n}$. (See Section IB for the channel function definition.) Solving for $\mathbf{a}$ from the first QCMT equation, $S_{\ell m}$ in Eq. (37) will be identical to the one in Eq. (24).

## III. APPLICATIONS OF THE MITTAG-LEFFLER EXPANSION

In this section we show the details of applying Mittag-Leffler expansion to

$$\psi_{\mathrm{scat}}(\mathbf{r}) = \mathrm{i}\omega\int \Gamma(\mathbf{r},\mathbf{r}',\omega)\Delta B(\mathbf{r}',\omega)\psi_{\mathrm{inc}}(\mathbf{r}'), \tag{38}$$

which leads to various QNM expansion formulae.

### A. Equivalence between decomposition approaches

#### 1. Green's function

If we use the full Green's function of the system and get $\psi_{\mathrm{scat}}$ from Eq. (38), rather than doing QNM expansions, we can identify the QNM expansion of Green's function

$$\Gamma(\mathbf{r},\mathbf{r}') = \Phi_{\mathrm{Rqnm}}(\mathbf{r})\frac{1}{\mathrm{i}(\Omega-\omega)}\frac{1}{N(\omega)}\Phi_{\mathrm{Lqnm}}^T(\mathbf{r}'). \tag{39}$$

This is different from the other more widely used form

$$\Gamma(\mathbf{r},\mathbf{r}') = \Phi_{\mathrm{Rqnm}}(\mathbf{r})\frac{1}{\mathrm{i}(\Omega-\omega)}\Phi_{\mathrm{Lqnm}}^T(\mathbf{r}'). \tag{40}$$

In deriving Eq. (39) we have used what is known as the orthogonality-decomposition approach, whereas the residue-decomposition approach is used to derive Eq. (40).

Note that at each resonant frequency $\widetilde{\omega}_i$, the $i$th row of the $N(\widetilde{\omega}_i)$ matrix is diagonal, and the non-zero element is responsible for the normalization of the $i$th mode. As a result, if we apply Mittag-Leffler to Eq. (39), we will obtain Eq. (40). This proves the equivalence between residue-docomposition approach to orthogonality-decomposition approach. This shows that the difference between Eq. (39) and Eq. (40) is because Mittag-Leffler expansion is applied to the same quality at different stages.

#### 2. QNM expansion formulae

If we apply Mittag–Leffler to $\Gamma$ in Eq. (40), we have $\Gamma(\mathbf{r},\mathbf{r}',\omega) = \sum_i \frac{\psi_{R,i}(\mathbf{r})\psi_{L,i}^T(\mathbf{r}')}{\mathrm{i}(\widetilde{\omega}_i-\omega)}$ and put it back to Eq. (38) and we obtain one version of the residue-decomposition approach. Similarly, we apply ML to $\omega\Gamma\Delta B$, and by the sum rules of $\Gamma$ and partial fractions, we have

$$\omega \Gamma(\boldsymbol{r}, \boldsymbol{r}', \omega) \sum_n \frac{\sigma_n}{\omega - \omega_n}$$

$$= \sum_{i,n} \frac{\psi_{R,i}(\boldsymbol{r})\psi_{L,i}^T(\boldsymbol{r}')}{\mathrm{i}(\widetilde{\omega}_i - \omega)} \frac{\widetilde{\omega}_i \sigma_n}{(\widetilde{\omega}_i - \omega_n)}$$

$$- \underbrace{\sum_{i,n} \frac{1}{\mathrm{i}} \psi_{R,i}(\boldsymbol{r})\psi_{L,i}^T(\boldsymbol{r}') \frac{\sigma_n}{(\widetilde{\omega}_i - \omega_n)}}_{=\Gamma(\omega_n)\sigma_n = 0}$$

$$+ \underbrace{\Gamma(\omega = \omega_n)}_{=0} \sum_n \frac{\omega_n \sigma_n}{(\omega - \omega_n)} + \underbrace{\omega \Gamma \sum_n \frac{\sigma_n}{\omega - \omega_n}\bigg|_{\omega=0}}_{=0}$$

$$= \sum_{i,n} \frac{\psi_{R,i}(\boldsymbol{r})\psi_{L,i}^T(\boldsymbol{r}')}{\mathrm{i}(\widetilde{\omega}_i - \omega)} \frac{\widetilde{\omega}_i \sigma_n}{(\widetilde{\omega}_i - \omega_n)}. \tag{41}$$

This gives the QNM expansion formula

$$a_i = \int \psi_{L,i}^T \left[ (B_\infty - B_\mathrm{b}) \frac{\omega}{\widetilde{\omega}_i - \omega} + \frac{\widetilde{\omega}_i}{\widetilde{\omega}_i - \omega} \sum_n \frac{\sigma_n}{\widetilde{\omega}_i - \omega_n} \right] \psi_\mathrm{inc}$$

$$= \int \psi_{L,i}^T \left[ (B_\infty - B_\mathrm{b}) + \frac{\widetilde{\omega}_i}{\widetilde{\omega}_i - \omega} \Delta B(\widetilde{\omega}_i) \right] \psi_\mathrm{inc}, \tag{42}$$

which is identical to the expression in Ref. [11] derived from an augmented Maxwell operator approach, and in Ref. [12] derived from residue expansion of Green's function together with more complex sum rules. Various other expansion expressions summaried in Ref. [1] can be proven to be equivalent, similar to pole methods above.

### B. Resonant and background part

In this section, we derive the expression of $H(\omega)$, the frequency dependent part of the background part. When applying Mittag-Leffler expansion to Eq. (5) (dropping $N(\omega)$), since elements in the scattering matrix are normally not bounded at complex infinity, the $H(\omega)$ term is generally inevitable, rendering the pole expansion formula of $S$ of no practical use. Although $H(\omega)$ cannot be obtained from Mittag-Leffler itself, by comparing

$$S = \underbrace{S_\mathrm{bg}(\omega = 0) + H(\omega) + \mathrm{i}\widetilde{K}\Omega^{-1}\widetilde{D}^T}_{\text{"background part"}} - \underbrace{\mathrm{i}\widetilde{K}(\Omega - \omega)^{-1}\widetilde{D}^T}_{\text{"resonant part"}}, \tag{43}$$

with Eq. (5), we conclude that

$$H(\omega) = S_\mathrm{bg}(\omega) - S_\mathrm{bg}(0) + \frac{\mathrm{i}\omega}{4\alpha\beta^*}(\Phi_\mathrm{inc}^\mathrm{TR}, \Delta B \Phi_\mathrm{inc}) + \widetilde{K}\frac{1}{\mathrm{i}\Omega}\widetilde{D}^T + K(\omega)\frac{1}{\mathrm{i}(\Omega - \omega)}D^T(\omega) - \widetilde{K}\frac{1}{\mathrm{i}(\Omega - \omega)}\widetilde{D}^T. \tag{44}$$

Due to cancellations of residues, the above expression is regular at each resonant frequency $\widetilde{\omega}_m$, hence an entire function. One can see that $H(\omega)$ contains the Born scattering term and effects of frequency-dependent coupling matrices.

---


\fi

\end{document}